\documentclass[review,3p]{elsarticle}
\usepackage{rotate,amssymb,amsmath,latexsym,epsf,epsfig,graphics,graphicx}
\usepackage[usenames, dvipsnames]{color}
\usepackage{verbatim,lineno,morefloats,soul,changes,textcomp,gensymb}
\usepackage[font=footnotesize,labelfont=bf]{subcaption}
\usepackage[bookmarks]{hyperref}
\modulolinenumbers[1]

\journal{Nuclear Instruments and Methods in Physics Research Section A}

\begin{document}
\begin{frontmatter}
\title{Performance in Test Beam of a Large-area and Light-weight GEM detector with 2D Stereo-Angle (U-V) Strip Readout.}
\tnotetext[mytitlenote]{This work is supported by Brookhaven National Laboratory through the eRD6 consortium within the EIC R\&D program.}
\author[1]{Kondo Gnanvo\corref{cor}}
\cortext[cor]{Corresponding author: Tel.: 1-434-924-6796; Email: \href{mailto:kgnanvo@virginia.edu}{\textbf{kgnanvo@virginia.edu}}.}
\author[1]{Xinzhan Bai}
\author[1]{Chao Gu}
\author[1]{Nilanga Liyanage}
\author[1]{Vladimir Nelyubin}
\author[2,3]{Yuxiang Zhao}
\address[1]{University of Virginia, Department of Physics, Charlottesville, VA 22904, USA}
\address[2]{Thomas Jefferson National Accelerator Facility, Newport News, VA 23606, USA}
\address[3]{University of Science and Technology of China, Hefei, Anhui 230026, China}
\begin{abstract}
A large-area and light-weight Gas Electron Multiplier (GEM) detector was built at the University of Virginia as a prototype for the detector R$\&$D program of the future Electron Ion Collider. The prototype has a trapezoidal geometry designed as a generic sector module  in a disk layer configuration of a forward tracker in collider detectors.  It is based on  light-weight material and narrow support frames in order to minimize multiple scattering and dead-to-sensitive area ratio. The chamber has a novel type of two dimensional (2D) stereo-angle readout board with U-V strips that provides (r,$\varphi$) position information in the cylindrical coordinate system of a collider environment. The prototype was tested at the Fermilab Test Beam Facility in October 2013 and the analysis of the test beam data demonstrates an excellent response uniformity of the large area chamber with an efficiency higher than  95\%. An angular  resolution of 60 $\mu$rad in the azimuthal direction and a position resolution better than 550 $\mu$m in the radial direction  were achieved with the U-V strip readout board. The results are discussed in this paper. 
\end{abstract}
\end{frontmatter}

\section{Introduction} \label{sec:Introduction}
Gas Electron Multiplier (GEM) detectors \cite{gem} are playing an increasing role in the instrumentation of high and medium energy particle physics experiments. A major breakthrough related to the fabrication of GEM foil over recent years, the single mask technique \cite{singleMask}, has opened the field for large-area and cost effective tracking detectors using GEM technology with proven performances such as a spatial resolution better than 70 $\mu$m, a rate capability higher than 2.5 MHz / cm$^2$ and high tolerance to radiation in strong background environment etc.  A one-meter-long, trapezoidal triple-GEM prototype, the EIC-FT-GEM chamber, was assembled at the University of Virginia (UVa), as part of the tracking and particle identification detector R$\&$D eRD6 program administered at the Brookhaven National Laboratory (BNL) for the future Electron Ion collider (EIC) \cite{eic}. The prototype is a light-weight chamber characterised by its low-mass ($\le$ 0.4\% $X_0$ radiaton length) material and narrow support frames to minimize the multiple scattering in order to reduce the conversion of background photons to electrons and to minimize dead-to-sensitive area ratio. It has a distinctive 2D stereo angle readout board with U-V strip configuration for the measurement of 2D coordinates in a cylindrical coordinate system. We tested the chamber in the high energy hadron beam of the Fermilab Test Beam Facility (FTBF) in October 2013, as part of the eRD6 sector test beam campaign. The performance of the chamber in test beam is presented in this paper. In Section \ref{sec:prototype}, we briefly describe the design of the prototype with an emphasis on the innovative fine pitch U-V strip 2D readout board. In Section \ref{sec:ftbf}, we describe the FTBF experimental setup with the large-area GEM and provide a brief introduction to the readout electronics and the data acquisition system. In Section \ref{sec:characterisation}, we evaluate the basic performances of the chamber and discuss various technical issues inherent to its large size and novel readout strips. Finally, Section \ref{sec:resolution} is dedicated to the detailed study of the spatial resolution of the chamber and the performances of U-V strip readout board. 

\section{Design and Construction of the large EIC-FT-GEM with 2D U-V strip readout} \label{sec:prototype}
The EIC-FT-GEM chamber has a  trapezoidal shape, designed as a sector module in the disk layer configuration of a forward tracker for future collider experiment such as the EIC. It has a new type of 2D readout consisting of two layers of long narrow strips in a stereo-angle (or U-V strip) configuration for position measurement of the incoming particles in the cylindrical coordinate system (r, $\varphi$). The detector has a standard triple-GEM design as the COMPASS GEM detectors \cite{compass}, which was also adopted for GEM trackers used in other experiments \cite{sbsgem,stargem,totem,muontomo}. It consists of a 3 mm ionization gap between the drift cathode and the top foil of a stack of three GEM foils which provides three stages of gas avalanche amplification from the charges of the primary ionization. The foils in the stack are separated by 2 mm transfer regions.  The induced charge from the drift of the electrons generated at the bottom of the last GEM foil are collected on the U-V strip flexible readout board, located 2 mm below the GEM stack. 
\subsection{GEM foil design}
The active area of the GEM foil is 100 cm long (length of the trapezoid) with the widths of the two parallel sides (at the inner radius and outer radius sides in a disk layer configuration) equal to 23 cm and 44 cm respectively. The 5 $\mu$m thick Cu electrode on the top side of the foil is segmented into 32 High Voltage (HV) sectors, each approximately 100 cm$^2$ in area, separated from adjacent sectors  by 100 $\mu$m wide strips along the radial direction. The connections of the HV sectors to the power supply are done through 2 $\times$16 electrodes located at the inner and outer radius sides of the foils. This design, by avoiding HV traces along the radial sides of the GEM, allows narrow GEM support frames to minimizes the dead-to-sensitive area ratio in a disk layer configuration. A picture of the GEM foil, fabricated at the CERN Printed Circuit Board (PCB) workshop, is shown in Fig. \ref{fig:eic_gemfoil}.
\subsection{Two dimensional U-V strip readout layer} \label{sec:readout}
\begin{figure}[!ht]
\vspace{-0.05cm}
\captionsetup{width=0.9\textwidth}
\centering
\begin{subfigure}[t]{0.485\textwidth}
\includegraphics[width=0.95\textwidth,natwidth=980,natheight=731]{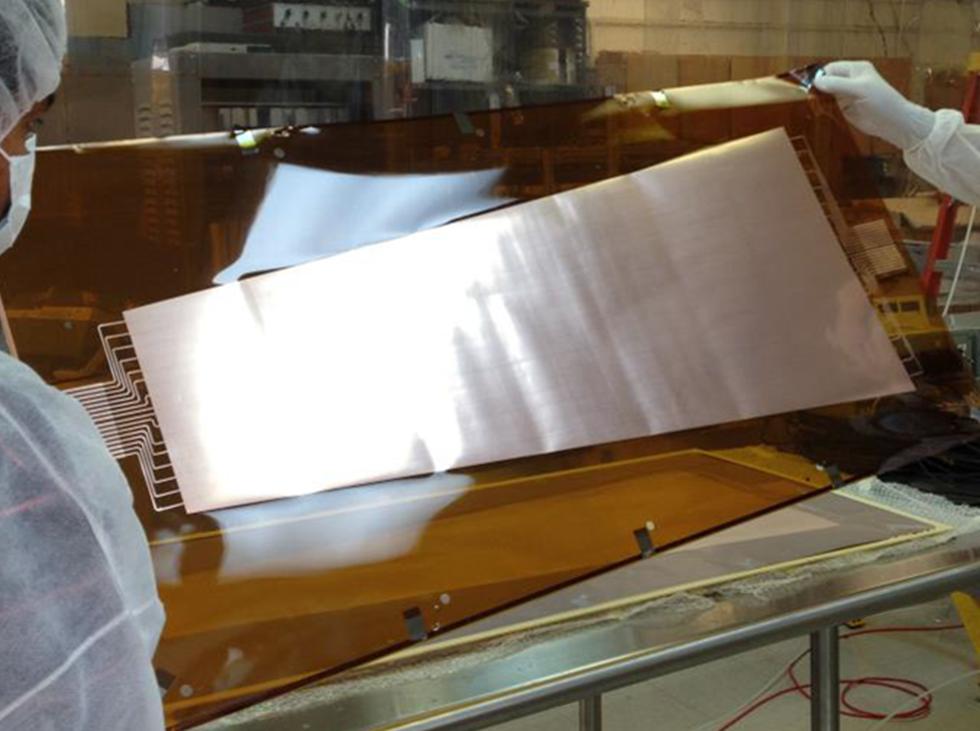}
\caption{\label{fig:eic_gemfoil} GEM foil}
\end{subfigure}
\hspace{-1em}
\begin{subfigure}[t]{0.485\textwidth}
\includegraphics[width=0.95\textwidth,natwidth=1500,natheight=1130]{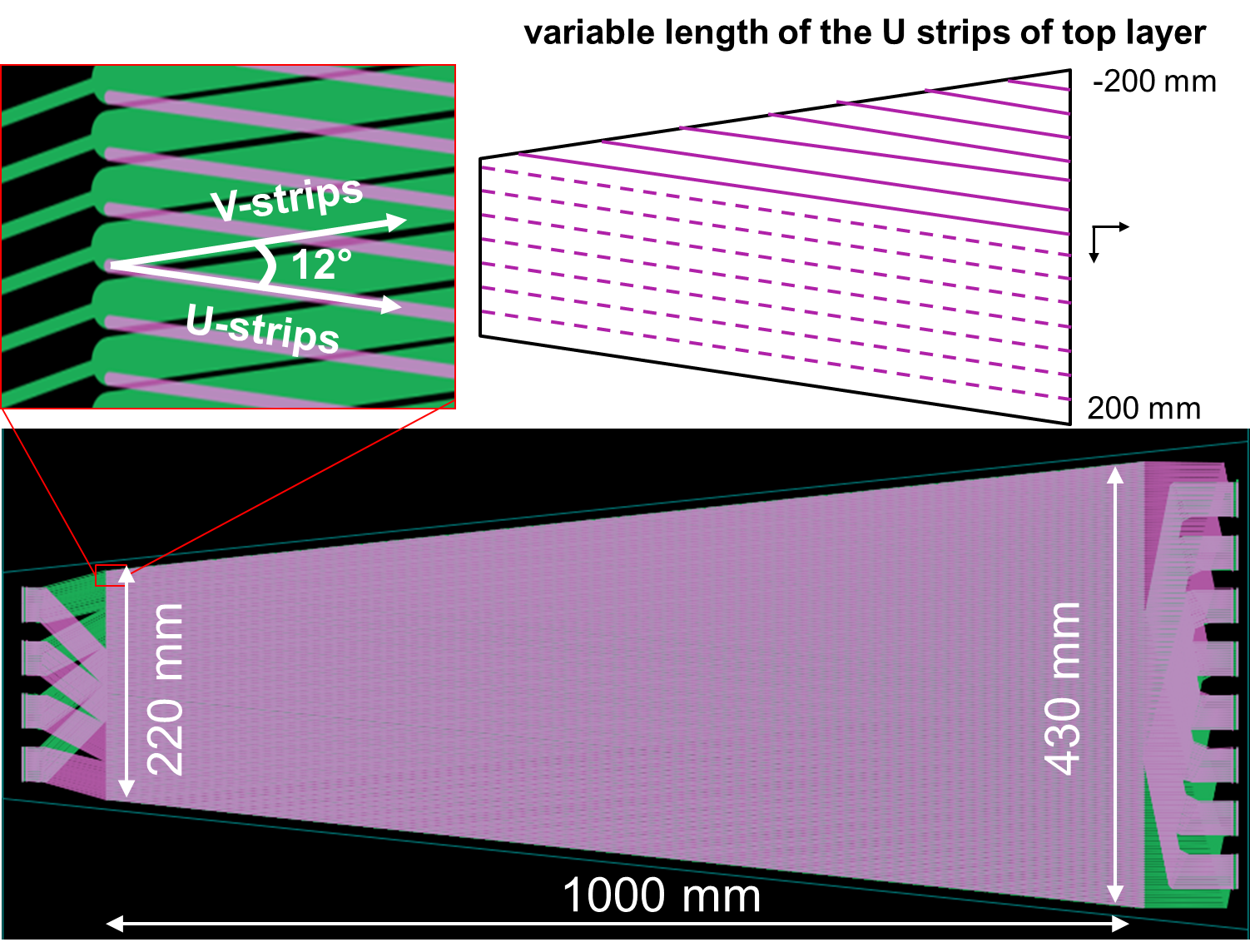}
\caption{\label{fig:eic_readout} 2D U-V strip readout}
\end{subfigure}
\caption{\label{fig:eic_design} EIC-FT-GEM chamber parts design}
\end{figure}

The readout board is a flexible double-layer printed circuit board (PCB) with an U-V strip pattern, etched on the same 50 $\mu$m thick copper-clad polymide material used to produce the GEM foils. The board is fabricated with the same technique used for the COMPASS 2D readout board \cite{compass,sbsgem}. The 140 $\mu$m wide U-strips of the top layer, run parallel to one of the two radial sides of the detector and the 490 $\mu$m wide V-strips of the bottom layer parallel to the other side. The different strip widths of top and bottom layers is to ensure an almost equal sharing of  charge coming  from the same event. The pitch is equal to  550 $\mu$m for both top and bottom layers and the angle between u and v strips is equal to 12\degree. With this design, only the two parallel sides of the chmber connect to the  Front End (FE) electronics. It is important, specially  in the disk layer configuration, to avoid having FE cards along the radial sides or at the back of the chamber in order to reduce the exposure of the electronics to high radiation as well as to minimize the material thickness in the sensitive area. Fig. \ref{fig:eic_readout} shows a sketch of the U-V strip readout board. Per design, half of the strips on each layer have a fixed length of approximately 1005 mm which is equal to the length of the radial side (see  dashed lines on cartoon at the top right of Fig. \ref{fig:eic_readout}) and the length of the other half set of strips decreases progressively as one moves towards the opposite side (straight lines on top right of Fig. \ref{fig:eic_readout}). The different strip lengths results in different pedestal noise level at different locations of the chamber which will be discussed later in Section \ref{sec:hvscanresults} of the paper. The flexible readout board was also fabricated at the CERN PCB workshop.
\subsection{Construction of the EIC chamber}
\begin{figure}[!ht]
\vspace{-0.cm}
\captionsetup{width=0.9\textwidth}
\centering
\includegraphics[width=0.9\textwidth,natwidth=1510,natheight=730]{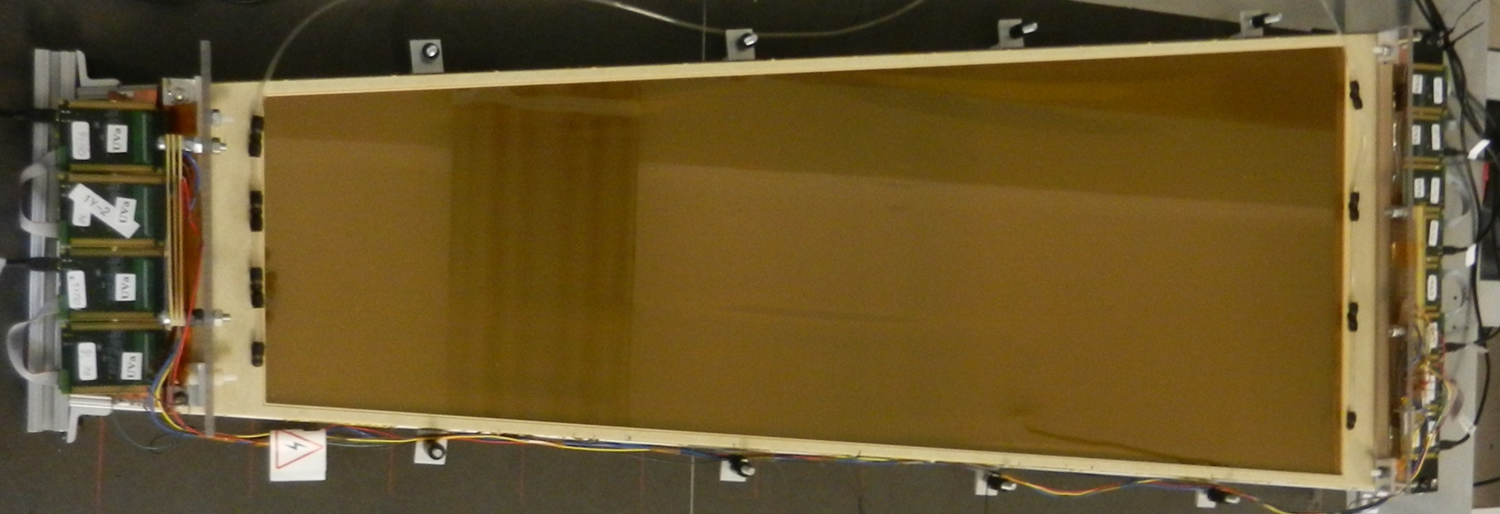}
\caption{\label{fig:eic_gem} EIC-FT-GEM chamber}
\end{figure}

Before the assembly of the chamber, each GEM foil as well as the drift cathode and the gas window foils are mechanically stretched and glued to dedicated support frames. The width of the frames are 8 mm for the radial sides of the trapezoid and 30 mm for parallel sides (inner and outer radius sides). The U-V strip readout foil is glued on a 3 mm thick Rohacell foam \footnote{\url{http://www.rohacell.com}} support which was found to be the best compromise for a rigid and low mass support structure. The stretched GEM foils stack is glued to the readout board, followed by the drift cathode foil which is basically a GEM foil with the Cu layer removed from the top side. Finally a 5 $\mu$m thick Kapton foil is glued on top of the  cathode foil to close the chamber and provide the gas entrance window. The assembly of the chamber, the resistive divider and the HV board design used to power the GEM foil HV sectors are similar to the one used for the GEM chambers of the Super Bigbite Spectrometer (SBS) \cite{sbsgem} at Jefferson Lab. A picture of the assembled chamber with its narrow radial frames and the FE electronic cards connected at the parallel sides is shown in Fig. \ref{fig:eic_gem}.
\section{Large EIC-FT-GEM chamber at the FTBF} \label{sec:ftbf}
The large  GEM chamber was tested at  the FTBF as part of the T-1037 slice sector test campaign, performed by the eRD6-FLYSUB in October 2013. The eRD6-FLYSUB is a consortium of researchers from Brookhaven National Laboratory (BNL), Florida Institute of Technology (FIT), Stony Brook University (SBU), University of Virginia (UVa) and Yale University, formed to carry out  detector R$\&$D for tracking and PID for EIC. In the consortium, UVa and FIT  focus on the development of new ideas for large-area GEM for EIC Forward Tracker (FT). The two groups shared the  MT62.b stand of the FTBF installing a total of  10 triple-GEM prototypes of various sizes and types in a common setup during the test beam campaign. The T-1037 effort was funded by the site-neutral EIC R$\&$D program administered at BNL.
\subsection{Experimental setup at the FTBF} \label{sec:ftbfsetup}
\begin{figure}[!ht]
\vspace{-0.cm}
\captionsetup{width=0.9\textwidth}
\centering
\begin{subfigure}[t]{0.485\textwidth}
\includegraphics[width=0.95\textwidth,natwidth=1425,natheight=1053]{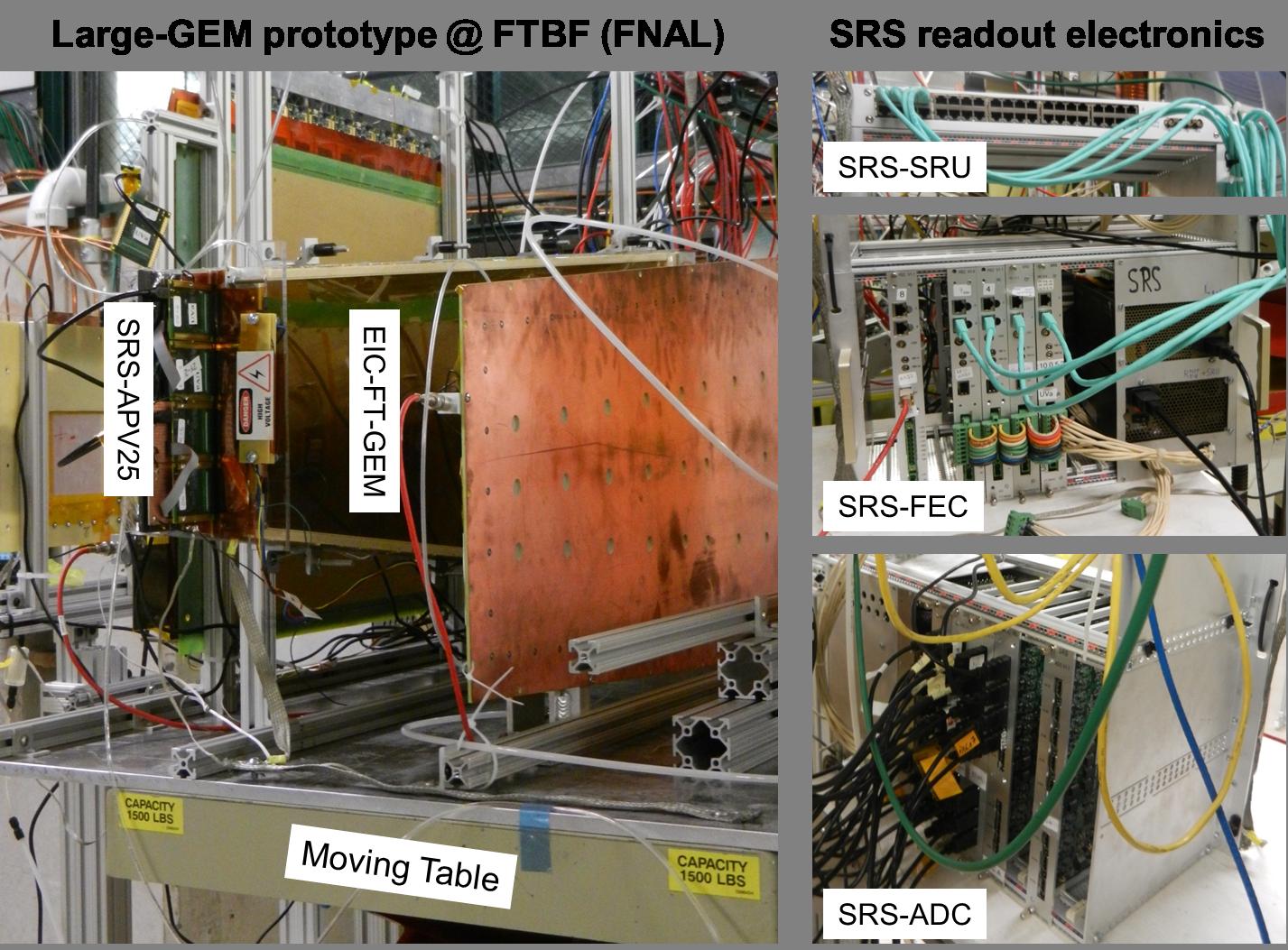}
\caption{\label{fig:ftbfEICSetup} EIC-FT-GEM chamber and SRS Electronics}
\end{subfigure}
\begin{subfigure}[t]{0.485\textwidth}
\includegraphics[width=0.95\textwidth,natwidth=1416,natheight=1103]{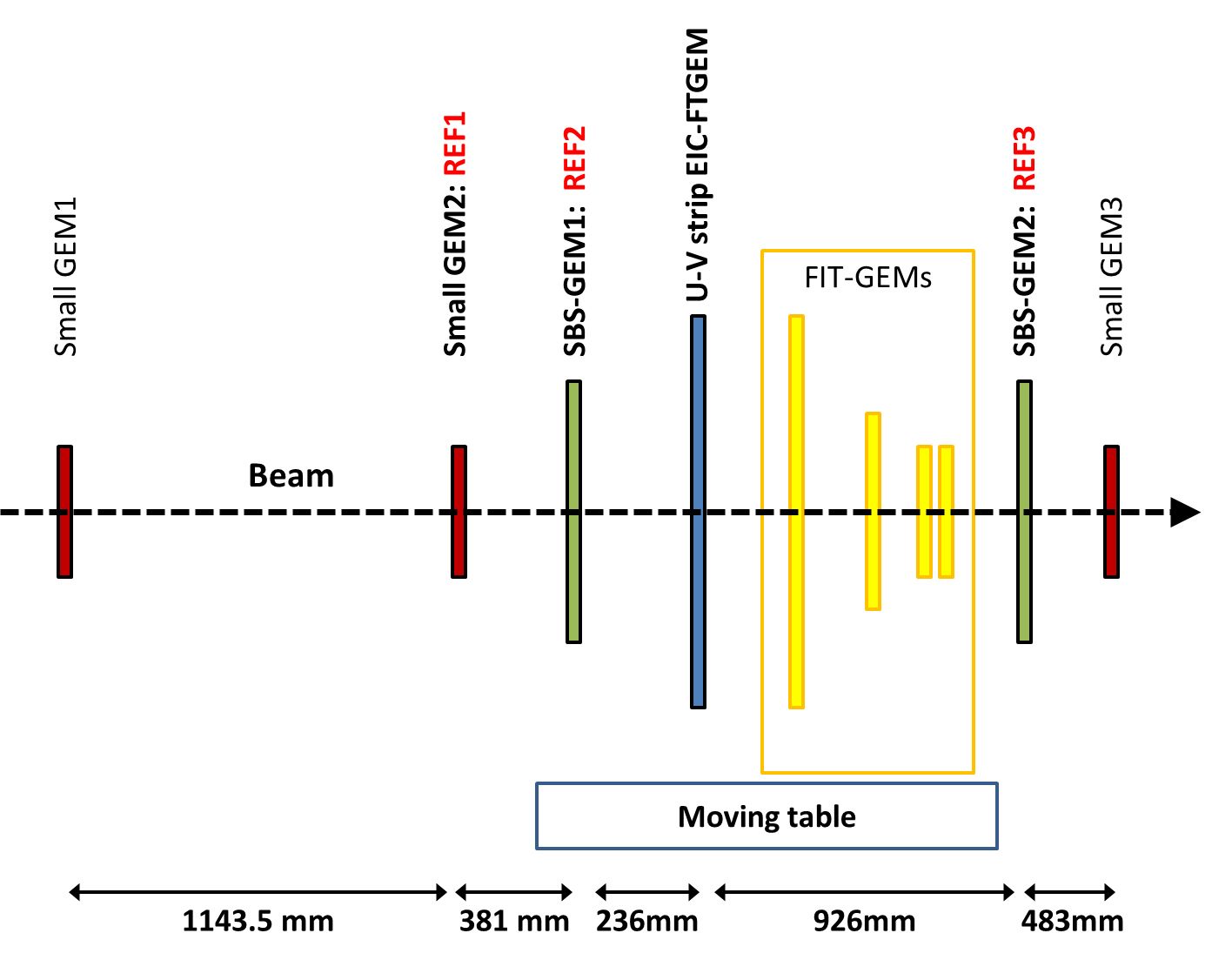}
\caption{\label{fig:ftbfLayout} Layout of 10 GEM chambers from UVa and FIT}
\end{subfigure}
\caption{\label{fig:ftbfSetup} Test beam setup of the large-area GEMs at the FTBF}
\end{figure}

The EIC-FT-GEM chamber was installed on the X-Y moving table along with a 50 $\times$50 cm$^2$ SBS GEM prototype \cite{sbsgem} and four other GEM chambers from FIT. In addition, three small (10 $\times$10 cm$^2$) GEMs and a second SBS GEM were installed at fixed locations in the beam line and used as reference trackers. The picture on the left of Fig. \ref{fig:ftbfEICSetup} shows a close view of the large trapezoidal EIC chamber. The layout with all 10 GEMs sitting on the stand is shown on the right. The facility provided high energy hadron beams of  120 GeV primary proton beam with the option for mixed hadron secondary beam of kaons, pions and protons at lower energy ranging from 20 GeV to 32 GeV. The beam has a cycle of 1 minute with 4 second spills of up to 2 $\times 10^4$ particles. Coincidence signal from plastic scintillators located upstream and downstream of the setup was also provided as external trigger signal for the readout electronics. 
\subsection{The Scalable Readout System Electronics at the FTBF}
The Scalable Readout System (SRS) \cite{srs} electronics based on the APV25 chip \cite{apv25} were used to readout all 10 GEM chambers during the test beam campaign. The SRS system, developed by the CERN-based RD51 international collaboration \cite{rd51collab}, is intended as a general purpose multi-channel readout solution suitable for a wide range of detector types, complexities and experimental environments. The scalability refers both to its wide range of applications and to its sizes varying from a few hundreds electronic channels to a large scale system with hundreds of thousands of channels in high energy particle experiments. Pictures of the different components of the SRS system used during the test beam at FTBF can be seen on Fig. \ref{fig:ftbfEICSetup}. A total of 8192 APV25 channels were used to read out all 10 GEM chambers. DATE \cite{alicedaq}  was used as SRS data acquisition software. A customized software package, the amoreSRS \cite{amoresrs}, built upon the AMORE framework \cite{alicedaq} and developed jointly by FIT and UVa, was used for both the online monitoring and the offline analysis of the test beam data. Both DATE software and the AMORE framework packages were developed for the ALICE experiment of the Large Hadron Collider (LHC) at CERN. 
\subsection{Computation of the pedestal data} \label{sec:pedestal}
\begin{figure}[!ht]
\vspace{-0.cm}
\captionsetup{width=0.9\textwidth}
\centering
\begin{subfigure}[t]{0.485\textwidth}
\includegraphics[width=0.95\textwidth,natwidth=1098,natheight=913]{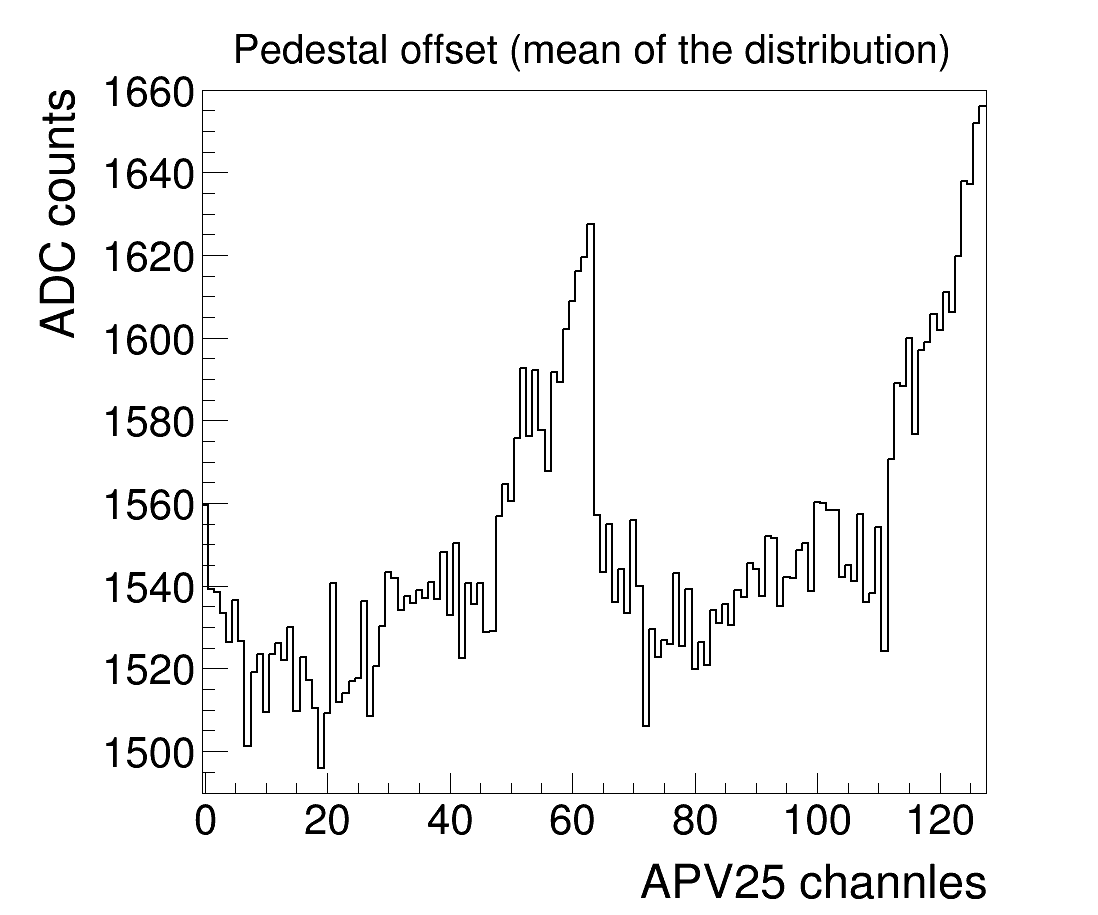}
\caption{\label{fig:eic_pedestaloffset} Pedestal offset (mean of the distribution)}
\end{subfigure}
\begin{subfigure}[t]{0.485\textwidth}
\includegraphics[width=0.95\textwidth,natwidth=1098,natheight=913]{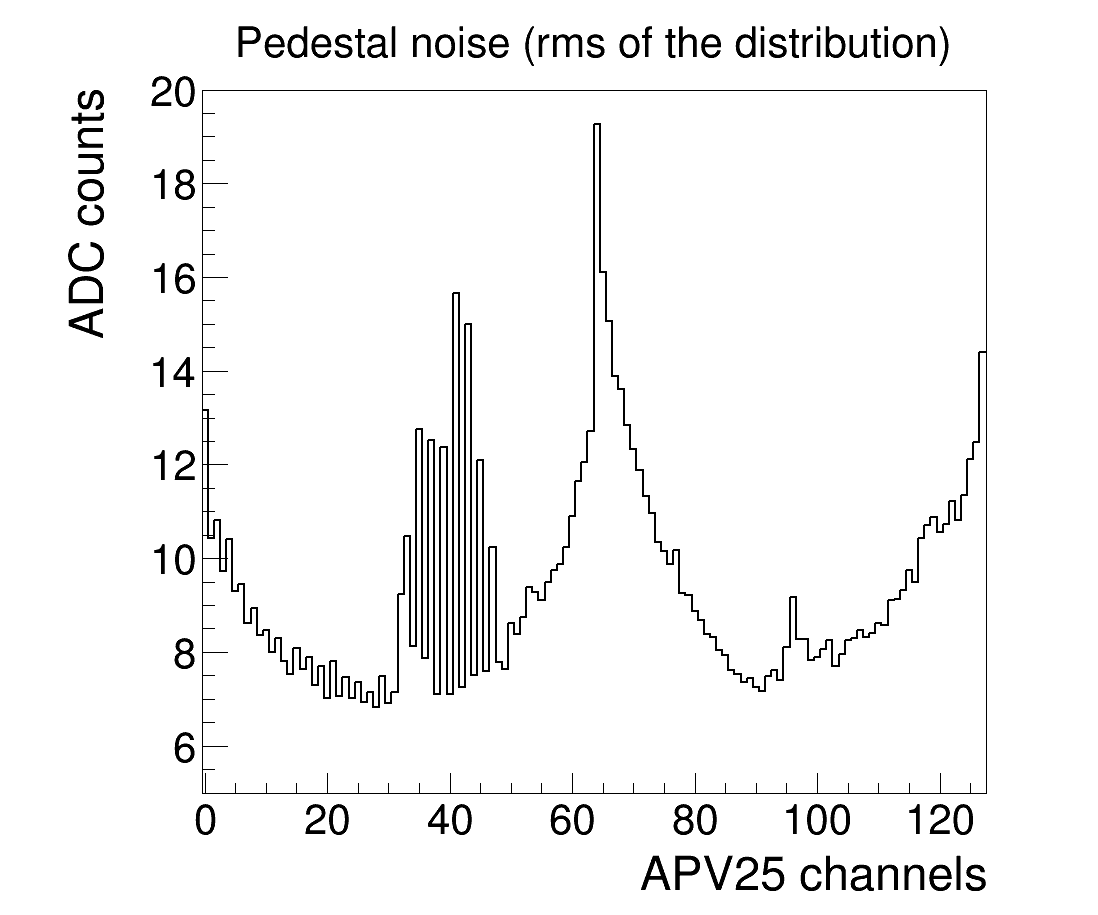}
\caption{\label{fig:eic_pedestalnoise} Pedestal noise (rms of the distribution)}
\end{subfigure}
\caption{\label{fig:eic_pedestal} Computed pedestal data for one of the APV25 FE cards (128 channels) connected to the readout board}
\end{figure}

Before each data run, a pedestal run with 2000 events is performed for all 10 GEM detectors. During the pedestal run, the voltage ($\Delta$V) across each GEM chamber  is maintained below 200V to prevent the ionization by cosmic particles. The average ADC counts is calculated and temporarily stored into histograms for each event and each time bin and for each APV25 channel connected to a readout strip. At the end of the pedestal run, the mean and the rms of each histogram, representing respectively the pedestal offset and noise of each channel, are saved into the output pedestal ROOT file. The ROOT file is loaded by the amoreSRS software for the online monitoring during the data acquisition as well as for the offline analysis of the test beam data. The zero suppression is  performed after common mode correction and pedestal offset subtraction by selecting the strips with an ADC counts above a threshold equal to n $\times$ $\sigma_{strip}$ as good hits. The parameter $\sigma_{strip}$ is the rms of pedestal associated to the given strip and n an integer representing the thresold parameter. For the results reported in this paper, the zero suppression threshold is set at 5 $\times$ $\sigma$ unless explicitly mentionned. The plots of Fig. \ref{fig:eic_pedestal} show the pedestal data for each of the 128 channels of one APV25 FE card connected to the chamber's readout board. The pedestal offsets, shown on Fig. \ref{fig:eic_pedestaloffset}, fluctuate around the baseline value of approximately 1550 ADC counts. The baseline is obtained after common mode correction performed individually for each APB25 FE cards using the pedestal data from all its channels. The pedestal offset values are used for the channel-by-channel pedestal subtraction before the zero suppression. The APV25 channel noise (rms of distribution) are shown on Fig. \ref{fig:eic_pedestalnoise} for the same APV25 FE card. From a previous study on the APV25 gain calibration \cite{apvgain},  one ADC count is estimated to be about 230 e- (or 0.037 fC). The pedestal noise varies from channel to channel with a minimum around 7 ADC counts (1610 e-) to a maximum of about 20 ADC counts (4600 e-) with more than 85\% of the channels haveing a noise level below 12 ADC counts (2760 e-) for this APV25 FE card. 
\begin{figure}[!ht]
\vspace{-0.cm}
\captionsetup{width=0.9\textwidth}
\centering
\begin{subfigure}[t]{0.485\textwidth}
\includegraphics[width=0.95\textwidth,natwidth=822,natheight=751]{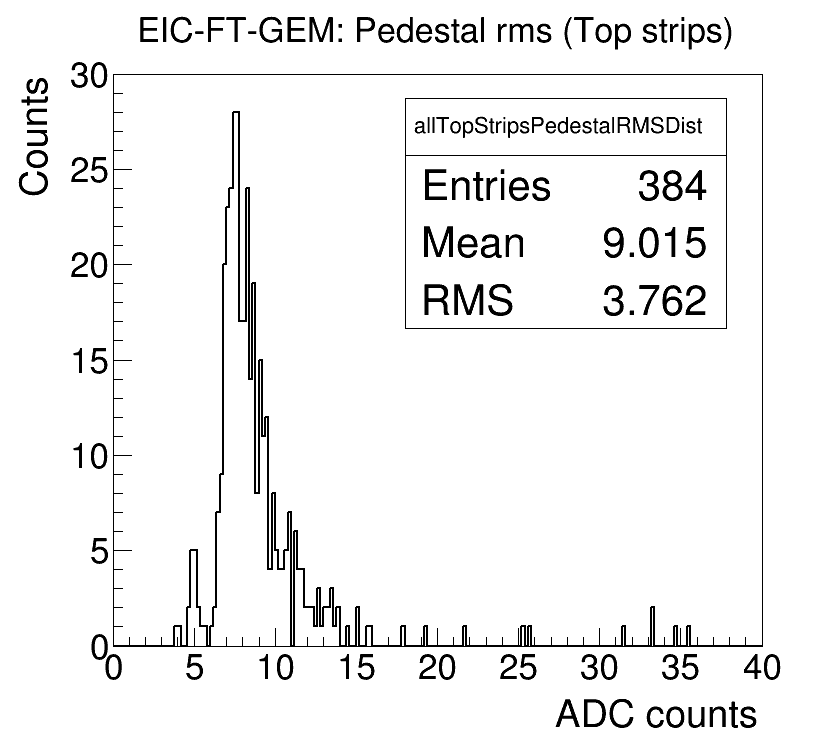}
\caption{\label{fig:eic_pedestaltoprmsdist} top layer}
\end{subfigure}
\begin{subfigure}[t]{0.485\textwidth}
\includegraphics[width=0.95\textwidth,natwidth=822,natheight=751]{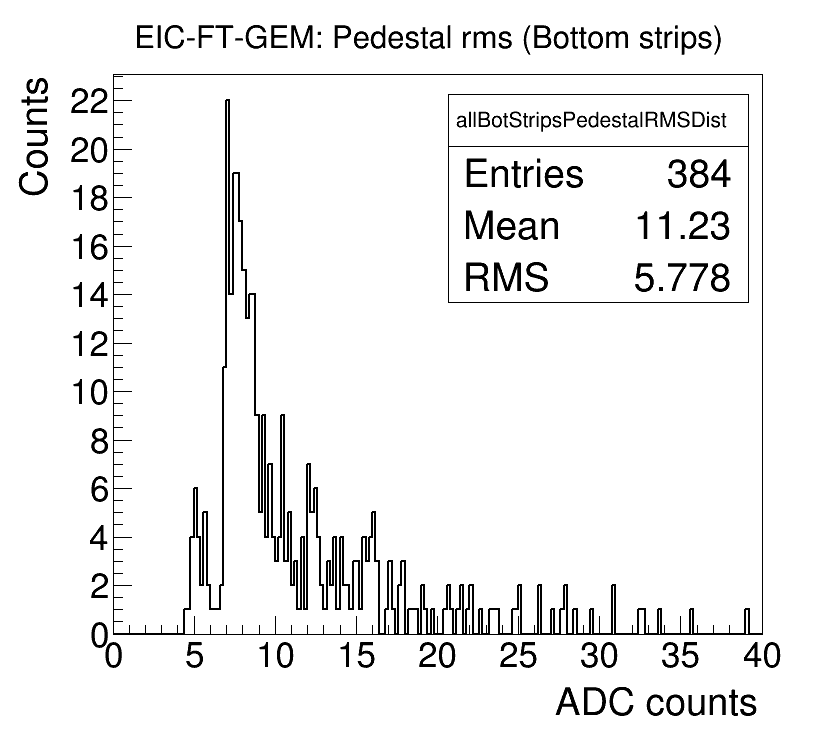}
\caption{\label{fig:eic_pedestalbotrmsdist}  bottom layer}
\end{subfigure}
\caption{\label{fig:eic_pedestalrmsdist} Distribution of the pedestal noise for APV25 FE channels connected to the chamber U-V strip readout}
\end{figure}
Per design, the strips of the bottom layer of the COMPASS-style 2D readout \cite{compass} are significantly wider than the strips of the top in order to achieve equal sharing of the cluster charges. Subsequently, the contribution of the strip capacitance noise to the overall pedestal noise is more pronounced for the bottom strips as shown on the plots of  Fig. \ref{fig:eic_pedestalrmsdist} for the distribution of the pedestal noise of 384 channels on each layer. The mean value of distribution of the pedestal noise for the bottom layer strips  is equal to 11.23 ADC counts (Fig. \ref{fig:eic_pedestalbotrmsdist}) which is about 25\% higher than the mean value for the top layer strips (9 ADC counts) shown on Fig. \ref{fig:eic_pedestaltoprmsdist}. 
\begin{figure}[!ht]
\vspace{-0.cm}
\captionsetup{width=0.9\textwidth}
\centering
\begin{subfigure}[t]{0.485\textwidth}
\includegraphics[width=0.95\textwidth,natwidth=1500,natheight=1099]{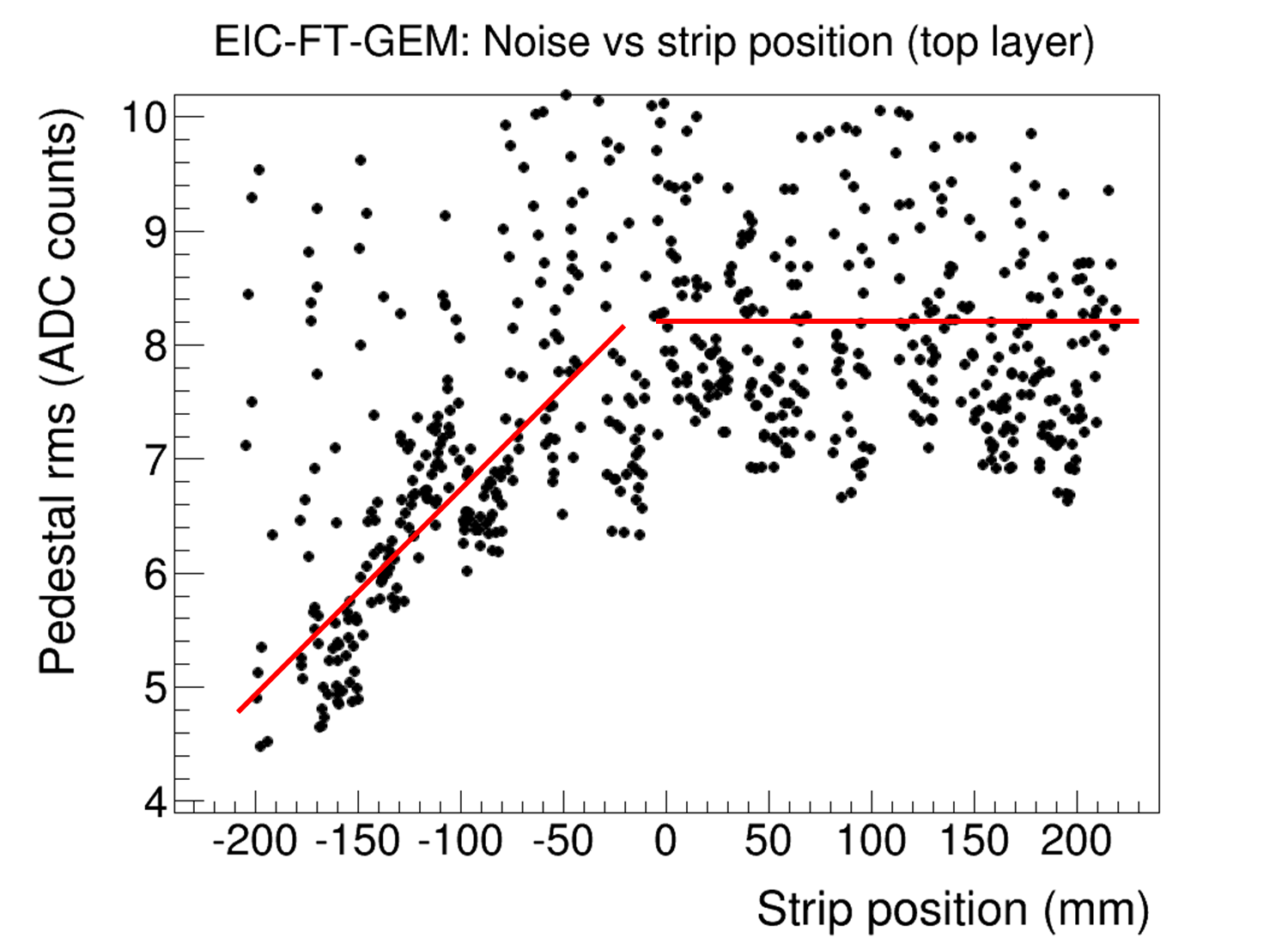}
\caption{\label{fig:eic_noise_topstrippos} top layer}
\end{subfigure}
\begin{subfigure}[t]{0.485\textwidth}
\includegraphics[width=0.95\textwidth,natwidth=1500,natheight=1099]{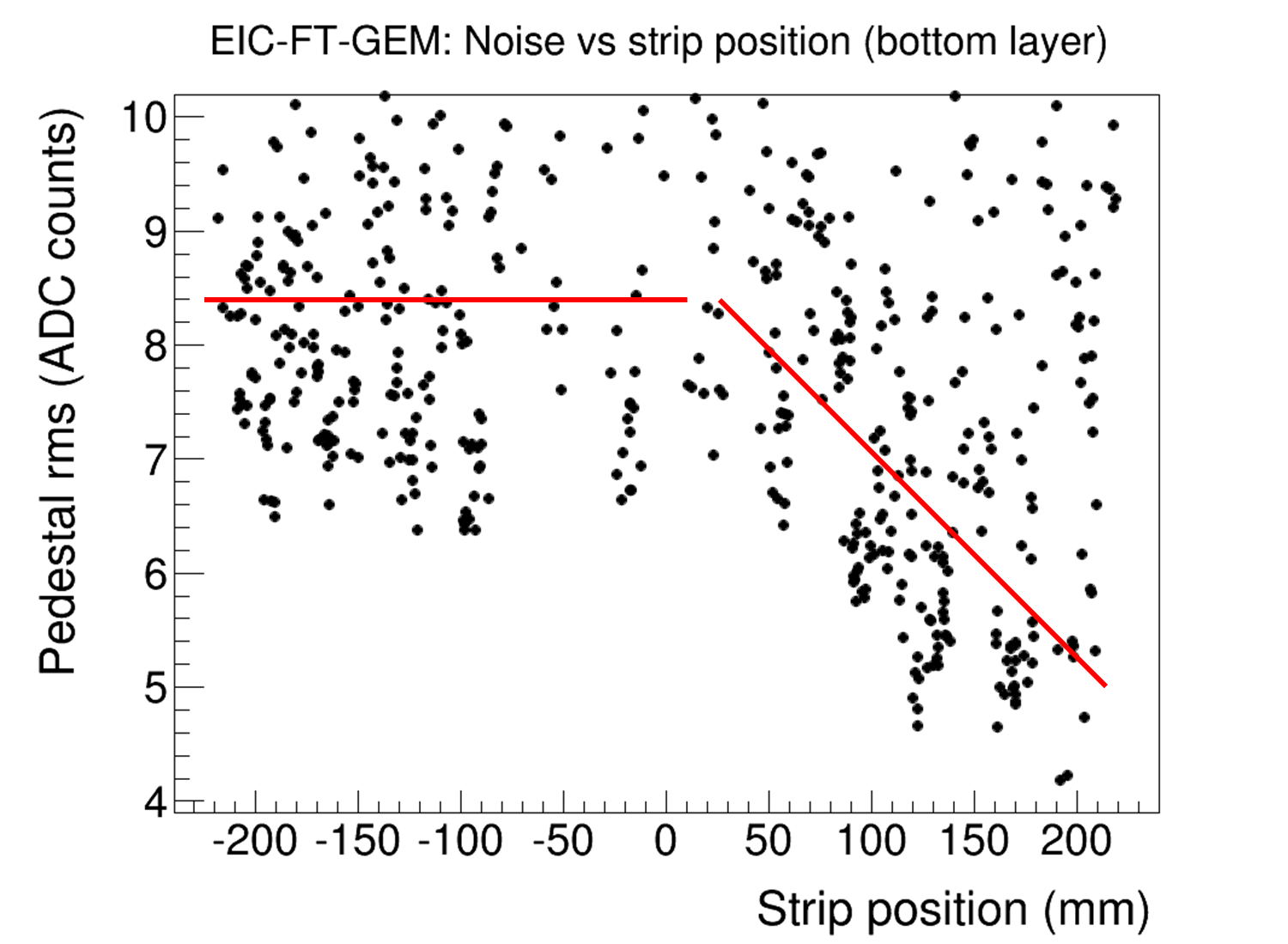}
\caption{\label{fig:eic_noise_botstrippos} bottom layer}
\end{subfigure}
\caption{\label{fig:eic_noise_strippos}  rms of the pedestal (noise level) vs. strip position on the outer radius side}
\end{figure}
\begin{figure}[!ht]
\vspace{-0.cm}
\captionsetup{width=0.9\textwidth}
\centering
\begin{subfigure}[t]{0.485\textwidth}
\includegraphics[width=0.95\textwidth,natwidth=1500,natheight=1099]{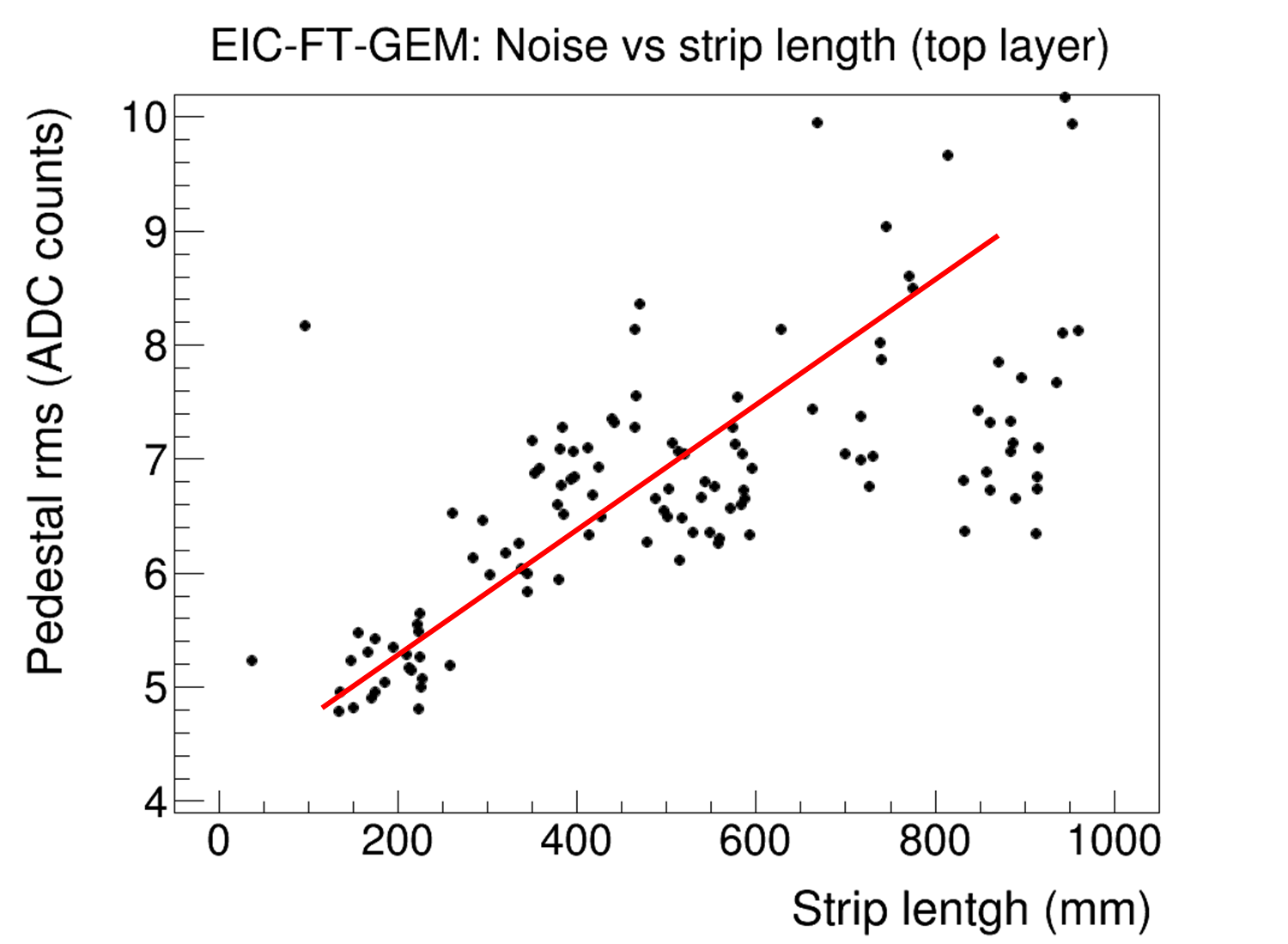}
\caption{\label{fig:eic_noise_topstriplength} top layer}
\end{subfigure}
\begin{subfigure}[t]{0.485\textwidth}
\includegraphics[width=0.95\textwidth,natwidth=1500,natheight=1099]{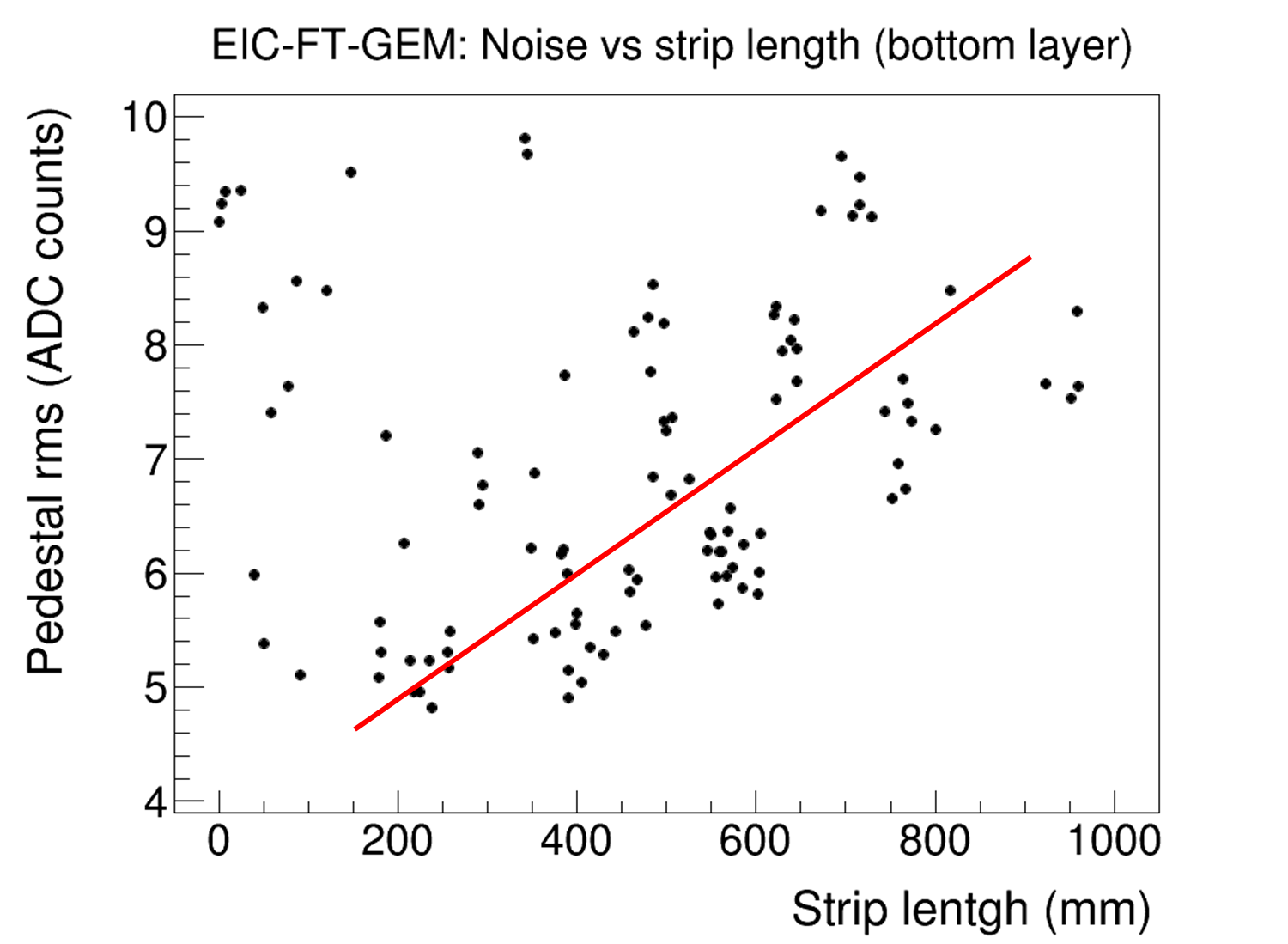}
\caption{\label{fig:eic_noise_botstriplength}  bottom layer}
\end{subfigure}
\caption{\label{fig:eic_noise_striplength}  rms of the pedestal (noise level) vs. strip length}
\end{figure}
Moreover, as we previously mentioned on Fig. \ref{fig:eic_readout}, Section \ref{sec:readout}, half of the strips for each layer (bottom or top) have a length that decreases progressively as one moves toward the radial side opposite to the strips direction. The impact of the different strip length is also seen on the pedestal noise of the APV25 channels. The plots on Fig. \ref{fig:eic_noise_strippos} show the pedestal noise as a function of the strip position in the chamber readout board.  Fig. \ref{fig:eic_noise_topstrippos} shows a somehow linear relationship between the strip position on the top layer and the average pedestal noise in the area of the chamber defined by the range of -215 mm to 0 mm on the outer radius side of the active area as illustrated on the top right sketch of Fig. \ref{fig:eic_readout}. This range corresponds to the area where the length of the strips varies with the position. The average noise level is constant on the other area of the chamber from 0 to 215 mm  corresponding to the constant strip length area. The distribution of the average noise level for the bottom layer strips follows an symmetrical pattern (constant from -215 mm to 0 and decreasing linearly from 0 to 215 mm) as shown on  Fig. \ref{fig:eic_noise_botstrippos}. This is expected since the strip design of the bottom layer is the mirror-image of the top layer strips. Fig. \ref{fig:eic_noise_striplength} shows the average noise level as a function of the strip length for both top (Fig. \ref{fig:eic_noise_topstriplength}) and bottom layer (Fig. \ref{fig:eic_noise_botstriplength}). The length of the strip is computed using the position of the strips and the geometrical parameters of the U-V strip readout board. The strips with constant length are not included in these results. Once again the correlation between average noise and strip length is clearly established.  
\section{Characterisation of the EIC-FT-GEM chamber} \label{sec:characterisation}
\subsection{Performances with the beam position scan} 
\begin{figure}[!ht]
\vspace{-0.cm}
\captionsetup{width=0.9\textwidth}
\centering
\begin{subfigure}[t]{0.485\textwidth}
\includegraphics[width=0.95\textwidth,natwidth=1344,natheight=711]{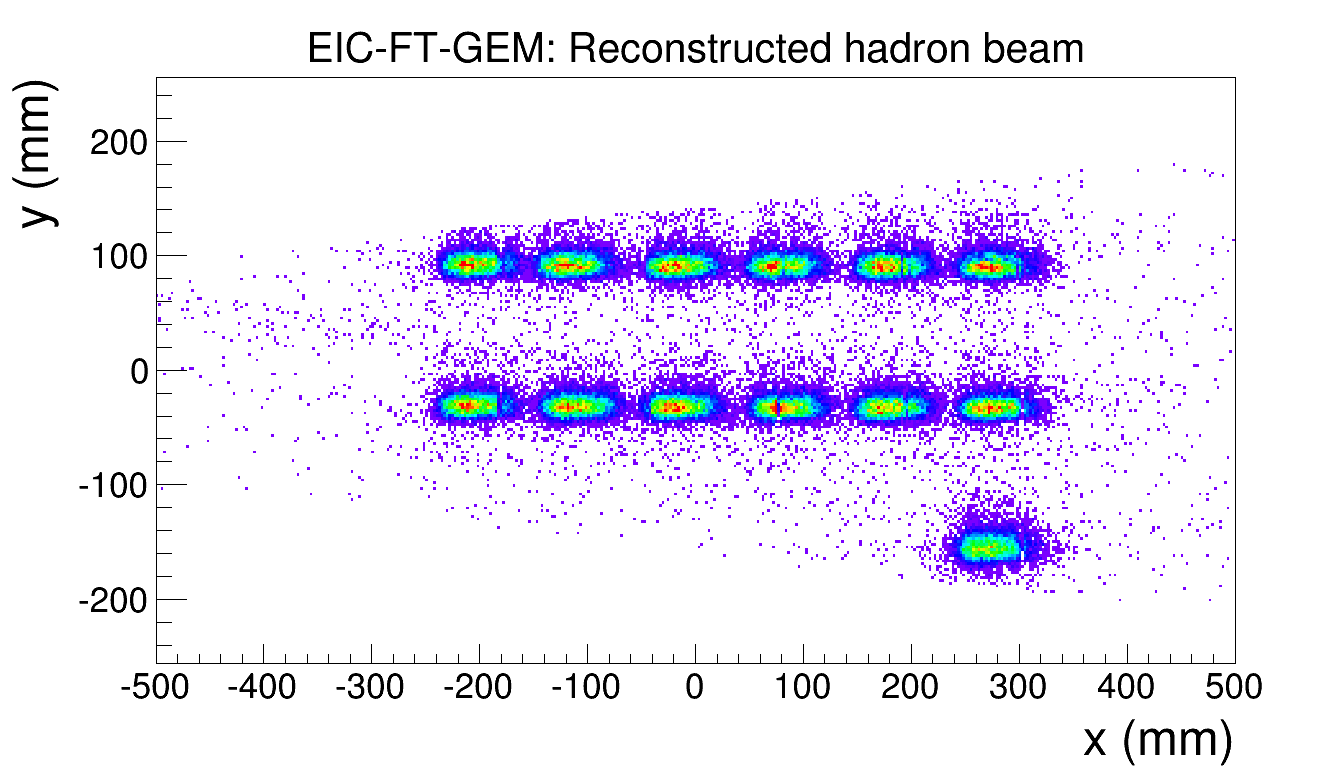}
\caption{\label{fig:eic_PositionScan} 2D beam profile from the reconstructed particle positions}
\end{subfigure}
\begin{subfigure}[t]{0.485\textwidth}
\includegraphics[width=0.95\textwidth,natwidth=1344,natheight=711]{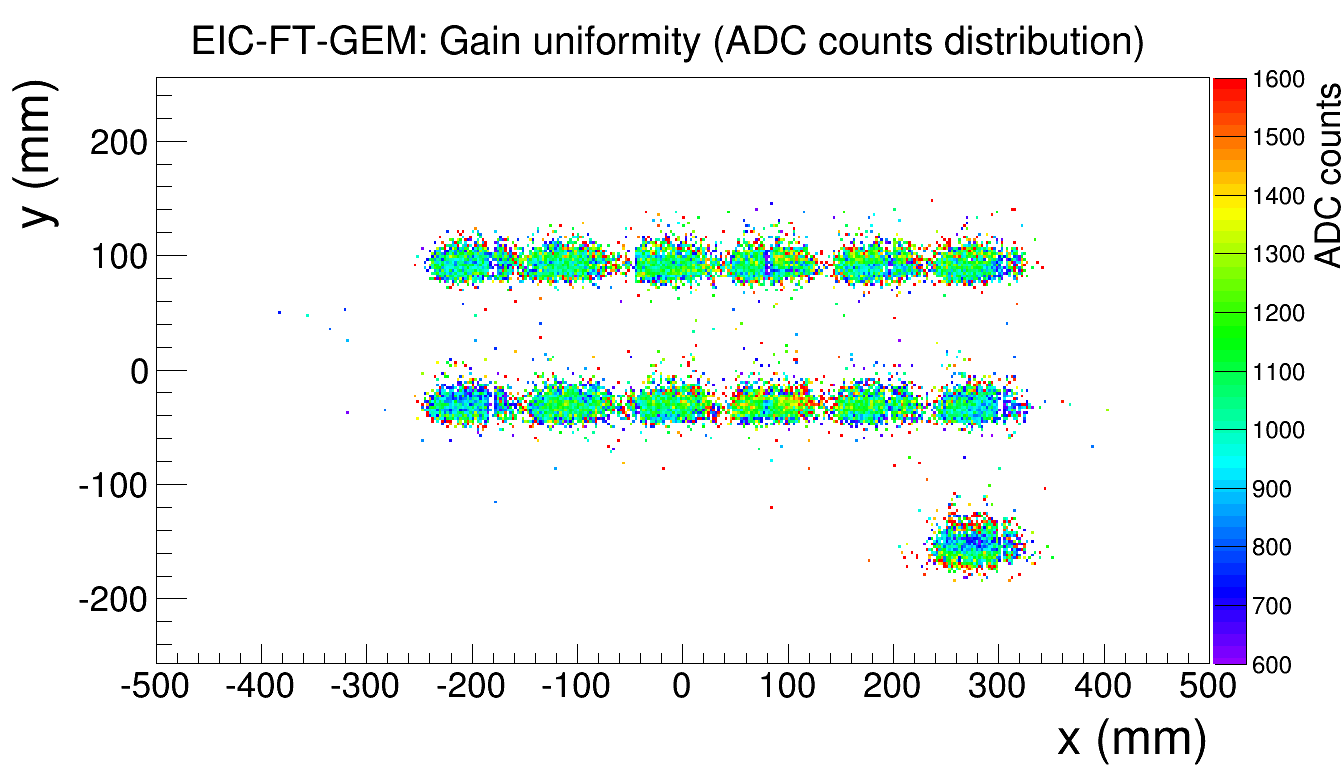}
\caption{\label{fig:eic_adcUniformity} Gain uniformity (average ADC counts per bin)}
\end{subfigure}
\caption{\label{fig:BeamScan} Position scan at 13 locations on the chamber with 32 GeV mixed hadron beam at FTBF }
\end{figure}
Position scan was performed with 32 GeV mixed hadrons beam to study the response uniformity of the EIC-FT-GEM chamber. Fig.\ref{fig:eic_PositionScan} shows the 2D beam profile from the reconstructed particle positions at 13 different locations on the chamber. The beam size is approximately  8 cm $\times$ 4 cm.  The coordinates ($x_{data}$, and $y_{data}$) of the hits in chamber are obtained from the Eq. \ref{eq:cartesianCoord}. Here the coordinates  $u_{data}$ and  $v_{data}$ which are measured along U and V direction respectively are calculated from the barycenter of the clusters on top and bottom strips. {\it L} is the length of the chamber and $\alpha$ the angle between U and V layer strips. 
\begin{align} \label{eq:cartesianCoord}
&x_{data} = 0.5 \times (L + (u_{data} - v_{data}) / \tan ( \alpha / 2 ) ) \nonumber\\
&y_{data} = 0.5 \times ( u_{data} + v_{data} ) 
\end{align}
Fig. \ref{fig:eic_adcUniformity} shows the 2D distribution of the average ADC counts for all 13 beam scan locations. The value of the average ADC counts at a given location is obtained by normalizing the ADC counts integrated over all events in each 2D bin by the number of entries in the bin. The  spatial distribution of the average ADC counts therfore represents the gain uniformity of the chamber. The gain variation observed from the test beam data is less than 20\% across the entire active area of the chamber.
\subsubsection{The spatial distribution of APV25 signal timing} \label{sec:gaspressure}
Gaseous detectors such as GEMs need to sustain a relatively high gas flow rate in order to avoid premature ageing when operating in huge background rate environment. At such high rate, the gas circulation structure of the detector need to be carefully designed to minimize pressure built up inside the chamber. However, for large area GEM chamber such as the EIC-FT-GEM chamber, one can not completely  eliminate the bending of the readout board caused by built-in over pressure and leading to a non-uniform induction gap  when light-weight material are used as readout support \cite{compass,sbsgem}.
\begin{figure}[!ht]
\vspace{-0cm}
\captionsetup{width=0.9\textwidth}
\centering
\begin{subfigure}[t]{0.6\textwidth}
\includegraphics[width=0.95\textwidth,natwidth=1500,natheight=860]{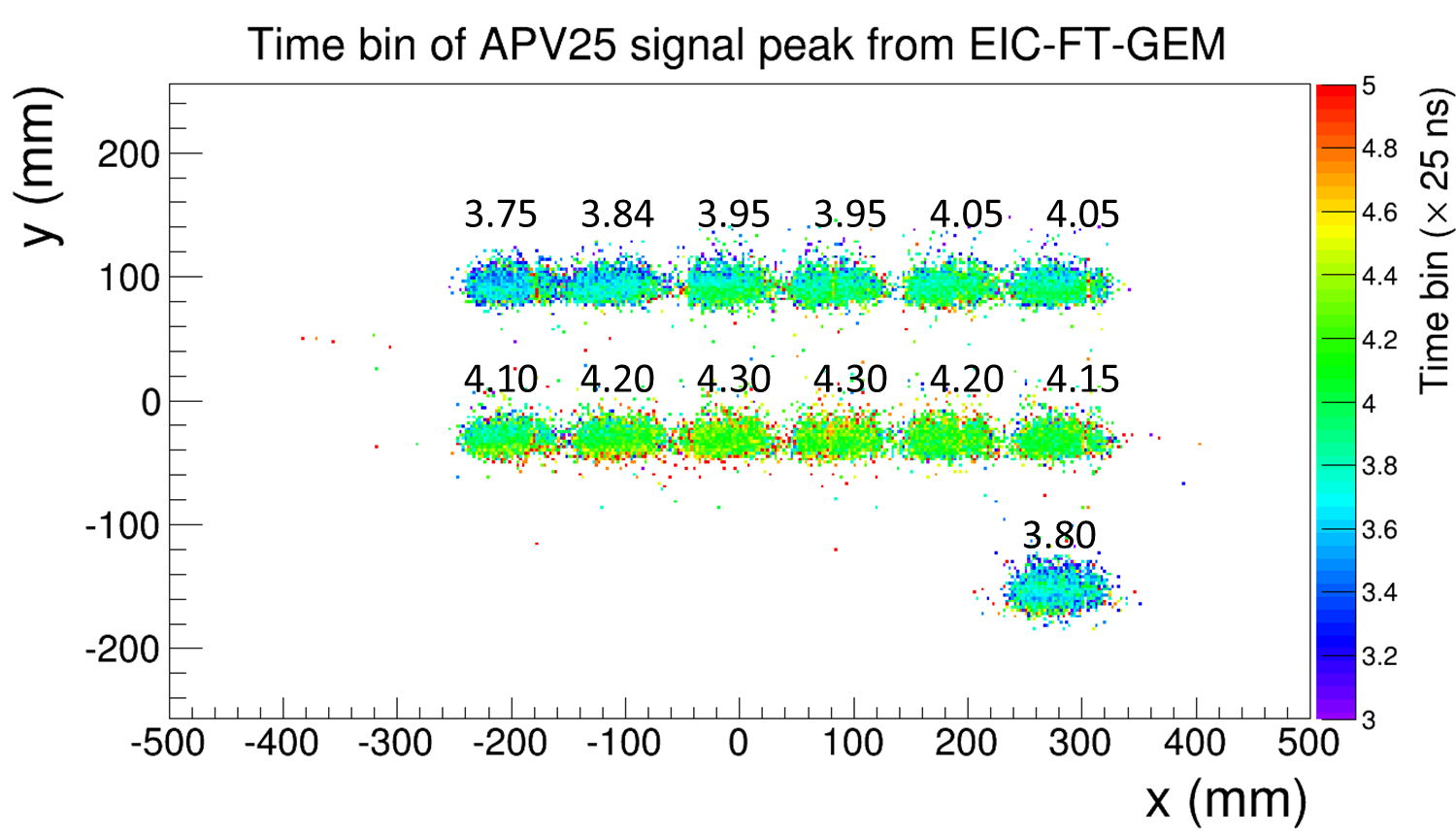}
\caption{\label{fig:eic_timing} spatial variation of the peak signal}
\end{subfigure}
\begin{subfigure}[t]{0.37\textwidth}
\includegraphics[width=0.95\textwidth,natwidth=822,natheight=791]{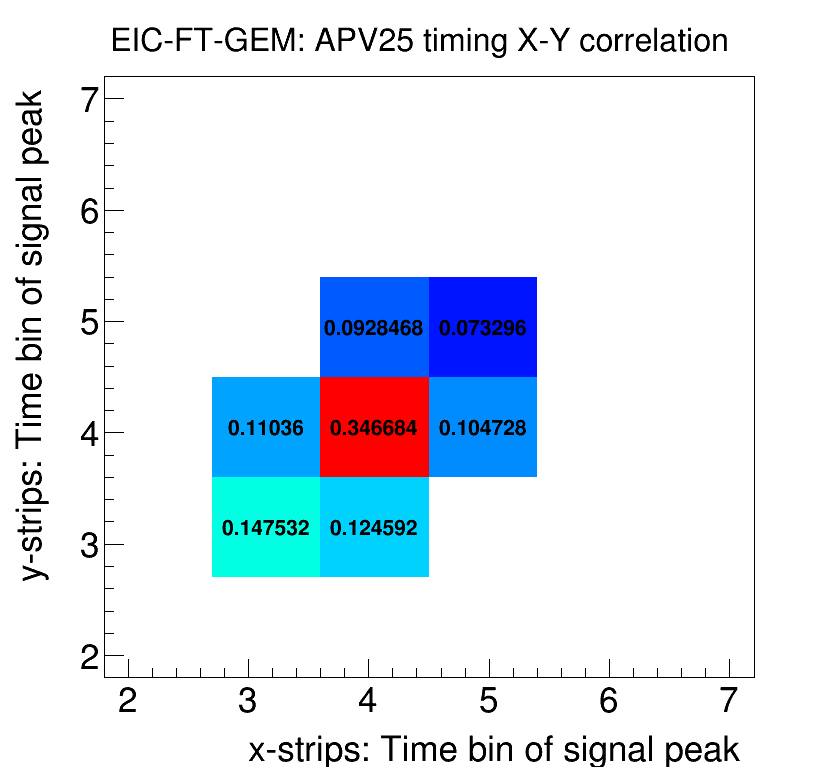}
\caption{\label{fig:eic_timingCorr} X-Y correlation of the peak signal}
\end{subfigure}
\caption{\label{fig:eic_signalPeak} Spatial variation of the APV25 peak signal in (time bin $\times$25 ns)}
\end{figure}
 We used the timing information provided by the APV25 signal as a probe to qualitatively evaluate the bending of the readout support. The delay of the signal peak, with respect to the trigger latency is defined by  the drift time of the charge in the induction region between the third GEM foil to the readout layer. For a given gas mixture, the drift time is a function of the gap in the induction region (drift length) and the electric field (drift velocity). Since the electric field in the induction region is inversely proportional to the induction gap, the drift time varies in quadrature with the gap between the third GEM foil and the readout board. The spatial distribution of the peak signal time from APV25 electronics is a simple way to qualitatively estimate the non-uniformity of the induction gap caused by the bending of the board. Fig. \ref{fig:eic_signalPeak} shows the 2D spatial distribution of the peak signal time for the strip with the largest ADC counts in the cluster. The signal peak timing is expressed in units of 25 ns time bin of the APV25 signal. The average time bin for all the events associated to a given position scan run  is shown on top of each reconstructed beam profile on Fig. \ref{fig:eic_signalPeak}. Higher time bin value means larger induction gap and lower electric field  and is an indication of local deviation of the readout from its ideal position. In Fig. \ref{fig:eic_timing}, the time bin value of the peak signal is slightly higher for the six positions at the center of the chamber (with the maximum equal to 4.30 [ $\times$ 25 ns]) and decreases to 3.75 [ $\times$ 25 ns] as one moves away from the middle of the chamber (see rows at -200 cm and 275 cm in Fig. \ref{fig:eic_timing}). The difference corresponds to a drift time variation of approximately 13.75 ns. The correlation plot of the time bin of the  signal peak for charge shared by top and bottom strips from the same event cluster is shown on Fig. \ref{fig:eic_timingCorr}. The percentage of the events falling in each bin is also displayed. All events show good correlation between signal time for x-clusters and y-clusters. Given the time resolution of APV25 the overall good correlation indicates the variation of APV25 peak signal time originates from the drift time of charge from the smae event shared by the strips of the two layers. Therefore, the spatial variation of the peak signal is a signature of the non uniform gap in the induction region caused by the bending of the readout support. This variation is significant and some further improvement in the gas flow design of the chamber is still necessary.
\subsection{Performances with the high voltage scan}  \label{sec:hvscanresults}
\subsubsection{Efficiency and relative gain} \label{sec:efficiency}
\begin{figure}[!ht]
\vspace{-0.0cm}
\captionsetup{width=0.9\textwidth}
\centering
\begin{subfigure}[t]{0.485\textwidth}
\includegraphics[width=0.95\textwidth,natwidth=1096,natheight=973]{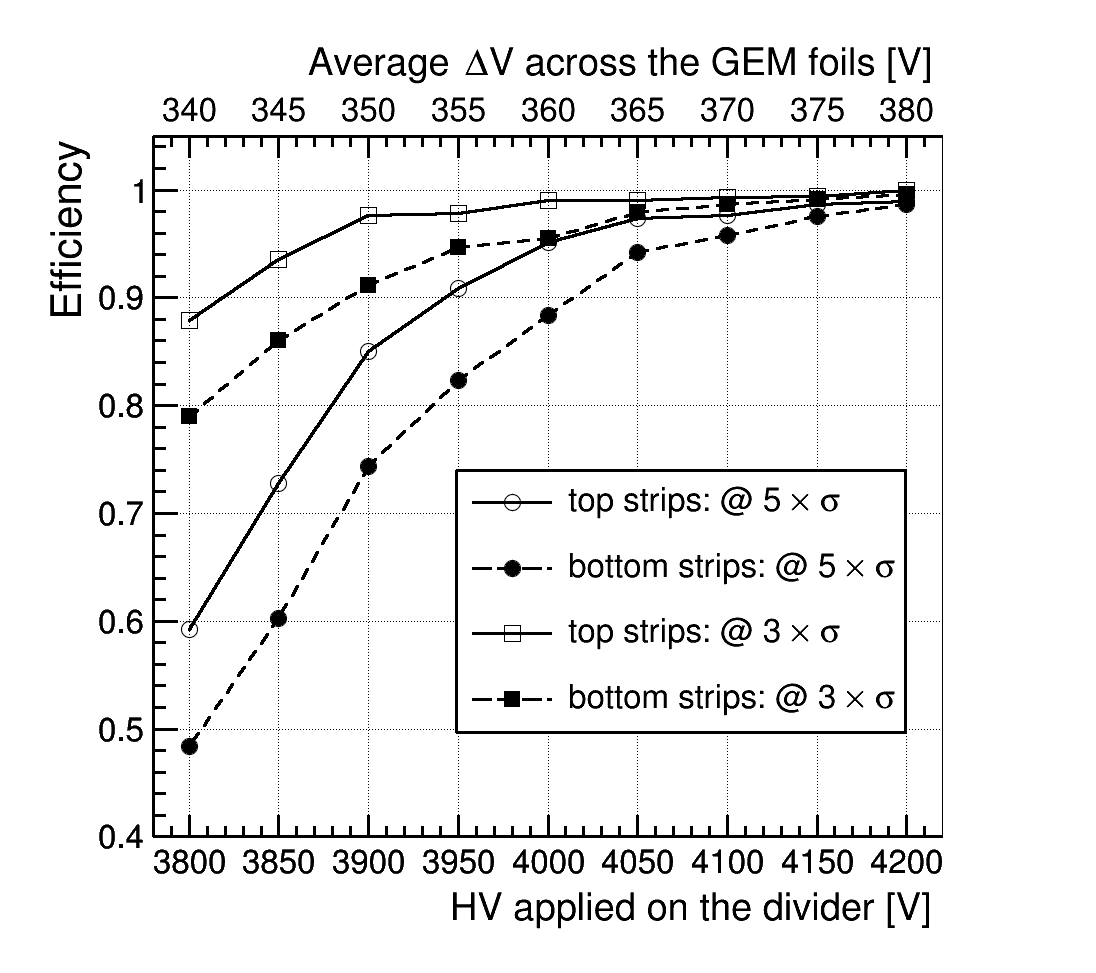}
\caption{\label{fig:eic_effvsHV} Efficiency at different threshold}
\end{subfigure}
\begin{subfigure}[t]{0.485\textwidth}
\includegraphics[width=0.95\textwidth,natwidth=1096,natheight=973]{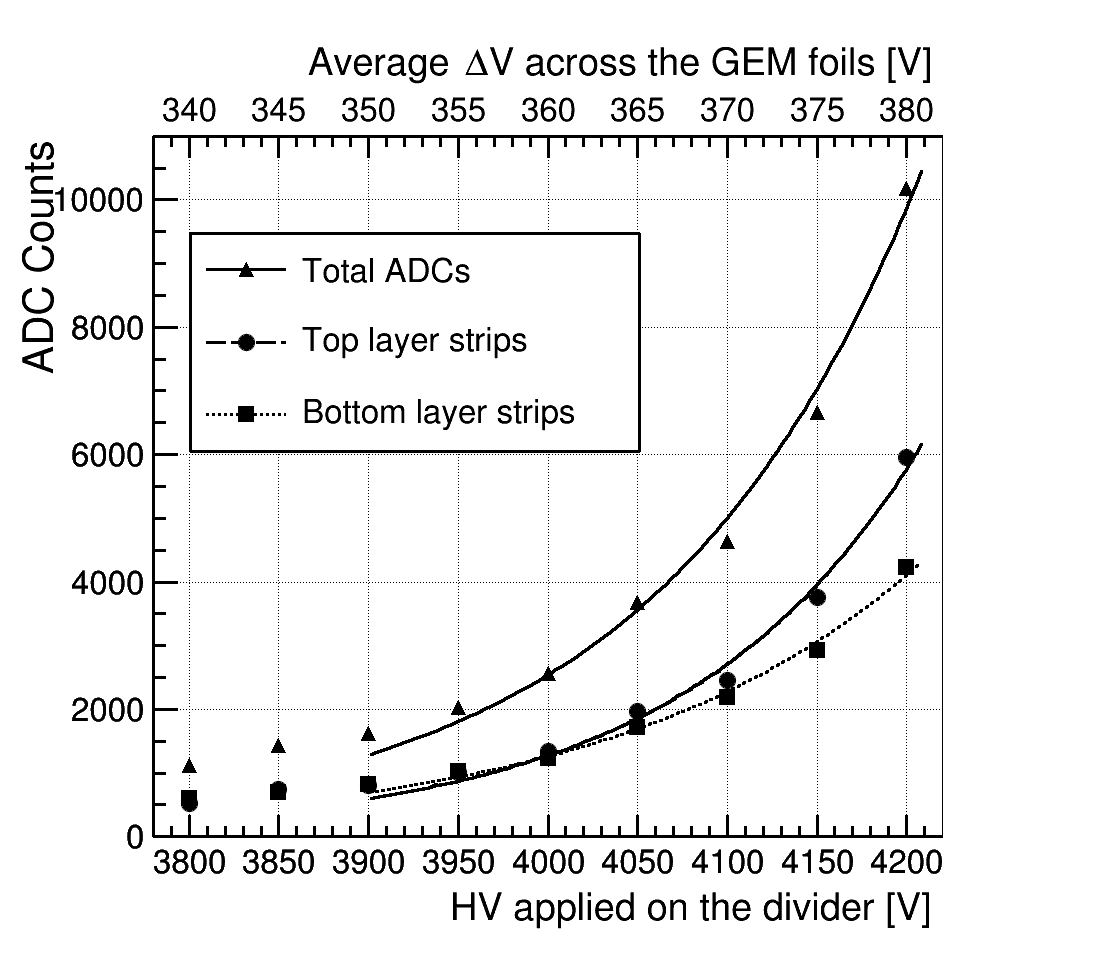}
\caption{\label{fig:eic_adcScan} Relative gain}
\end{subfigure}
\caption{\label{fig:HVScan} Efficiency and relative gain curve as a function of the high voltage}
\end{figure}

A high voltage (HV) scan was performed to study the gain variation and detector efficiency. For each HV scan run of 10,000 events, the voltage across the resistive divider was increased by steps of 50 V, starting from 3800 V up to 4200 V. A good event is defined as an event with at least one cluster which contains two or more consecutive strips with ADC counts above the zero suppression threshold. The efficiency is obtained  individually for top and bottom readout planes by normalizing the number of good events on each plane to the total number of ionizing tracks through the chamber. A valid track is required to have at least one cluster in both planes (x and y) of the 3 reference trackers REF1, REF2, REF3, (see Fig. \ref{fig:ftbfLayout}). Fig. \ref{fig:eic_effvsHV} shows the efficiency curves as a function of the applied voltage for two different thresholds, 3 $\times$ $\sigma$  and 5 $\times$ $\sigma$. For the 3  $\times$ $\sigma$ the efficiency is higher than 90\% for both top and bottom strips  even at a relatively low voltage of 3900 V on the divider, (average of 355 V across the GEM foils). High efficiency at low gain may be an indication of false positive events with noisy channels and bad hits being selected at  3 $\times$ $\sigma$. In order to have clean data for the analysis, a 5 $\times$ $\sigma$ threshold was found to be the best compromise for an effective suppression of  most false positive events while maintaining a high efficiency above 95\% for both top and bottom layers as observed for voltage 4100 V on the divider (373 V across the  GEMs). Lower efficiency is expected for the bottom layer strips as observed on the plots because of larger pedestal noise level caused by wider strip  as described in Section \ref{sec:pedestal}. 
The average ADC counts plots are shown  on Fig. \ref{fig:eic_adcScan} as a function of the voltage on the divider (HV) for top and bottom layer strips (circle and square markers respectively) as well as the total ADC counts (triangle markers). For each HV run, the average charge is derived from the most probable value (MPV) of the fit to Landau distribution expected for the energy deposit distribution of Minimum Ionising Particle (MIP).  Above 3900 V, average ADCs  fit nicely with the typical exponential behavior of the gain curve of a triple-GEM detector. 
\subsubsection{Cluster size and charge sharing} \label{sec:clusterSize}
\begin{figure}[!ht]
\vspace{-0cm}
\captionsetup{width=0.9\textwidth}
\centering
\begin{subfigure}[t]{0.485\textwidth}
\includegraphics[width=0.95\textwidth,natwidth=1096,natheight=973]{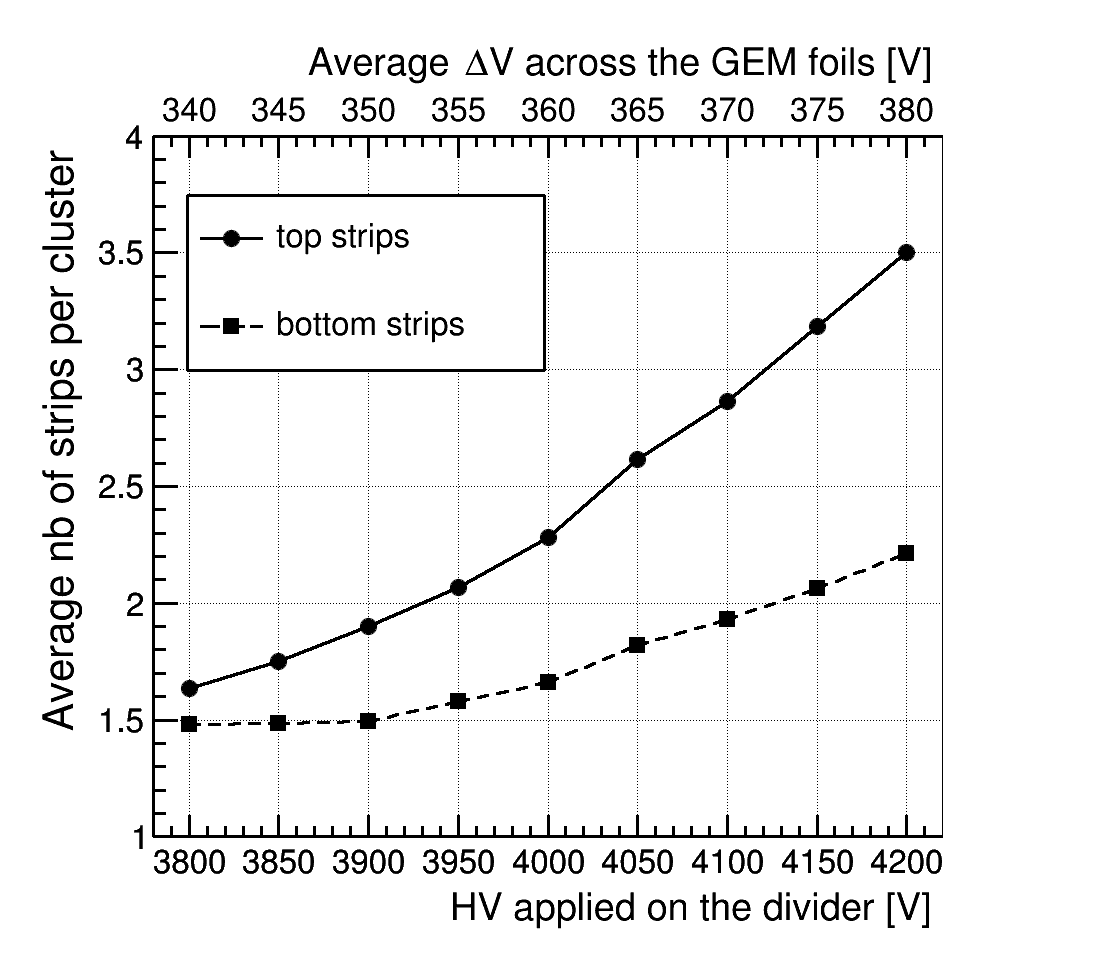}
\caption{\label{fig:clusterSize} cluster size}
\end{subfigure}
\begin{subfigure}[t]{0.485\textwidth}
\includegraphics[width=0.95\textwidth,natwidth=1096,natheight=973]{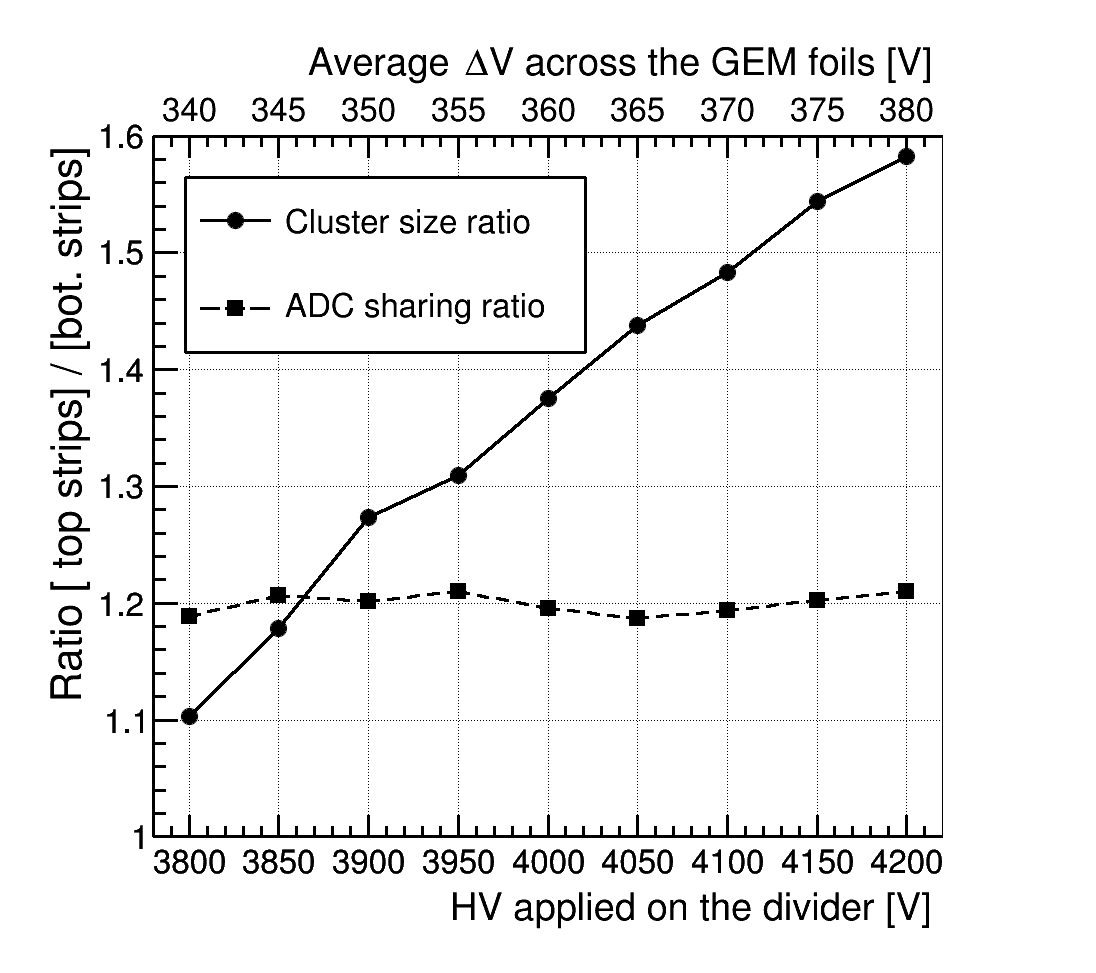}
\caption{\label{fig:chargeSharing} charge sharing and cluster charge ratio}
\end{subfigure}
\caption{\label{fig:chargeSharingAndClusterSize} Performance of the 2D U-V strip readout board}
\end{figure}

The  cluster size,  defined as the average number of consecutive strips with hits (ADC counts over threshold), is shown in Fig. \ref{fig:clusterSize}. The cluster size for the top layer strips (circle markers) increases steadily with increasing high voltage from 1.6 to 3.5  while the increase is less pronounced for the bottom layer strips (square markers) with only from 1.5 to 2.2 in the same range of applied voltage. 
The cluster size ratio between top and bottom strips is shown in Fig. \ref{fig:chargeSharing} (circle markers), and varies from 1.1 to 1.6 as a function of the applied voltage, however, the charge sharing ratio (square markers) is constant at 1.2 of the applied voltage. These two observations lead to the conclusion that most charge is collected by the two central strips of the cluster when the chamber is operated at a voltage higher than 4000 V. For the top strips, with lower pedestal noise, the charge collected by neighboring strips are enough to pass the  5 $\times$ $\sigma$ threshold, these strips however carry only a small fraction of total charges and therefore contribute little to the total cluster charges. In contrast, for the bottom strips,  higher pedestal noise  (see Section \ref{sec:pedestal}) means that many more neighboring strips in the tails of the cluster charge distribution are cut out by the zero suppression and only the two central strips constitute the cluster. 
\section{Spatial resolution studies of the EIC-FT-GEM chamber} \label{sec:resolution}
We report in this section the spatial resolution performance of the large chamber with the 2D U-V strip readout board. The study  was done with the data from 120 GeV primary proton beam in order to minimize the impact of the multiple Coulomb scattering and the track fit error on the precision of the measured resolution. The 2D proton beam profiles from the reconstructed particles are shown in Fig. \ref{fig:eic_protonBeamScan} at six different locations on the chamber. The number  associated to each position scan run is shown  on top of  each reconstructed beam spot on the plot. The spatial resolution, measured for the top and bottom strips along u and v directions, as well as in Cartesian (x,y) and cylindrical (r, $\varphi$) coordinates are reported in the following sections. 
\subsection{Alignment parameters}
Prior to the resolution analysis, a few correction steps are performed to produce the alignment parameters to compensate for misalignment of the reference trackers and the large chamber. The first of these steps is the offset correction to account for the relative shift of the origins in x and y directions. This is followed by the x-y plane rotation correction. All alignment constants are calculated relative to the reference tracker REF1.
\subsubsection{Correction of the offset}
\begin{figure}[!ht]
\vspace{-0.50cm}
\captionsetup{width=0.9\textwidth}
\centering
\begin{subfigure}[t]{0.625\textwidth}
\includegraphics[width=0.95\textwidth,natwidth=1500,natheight=866]{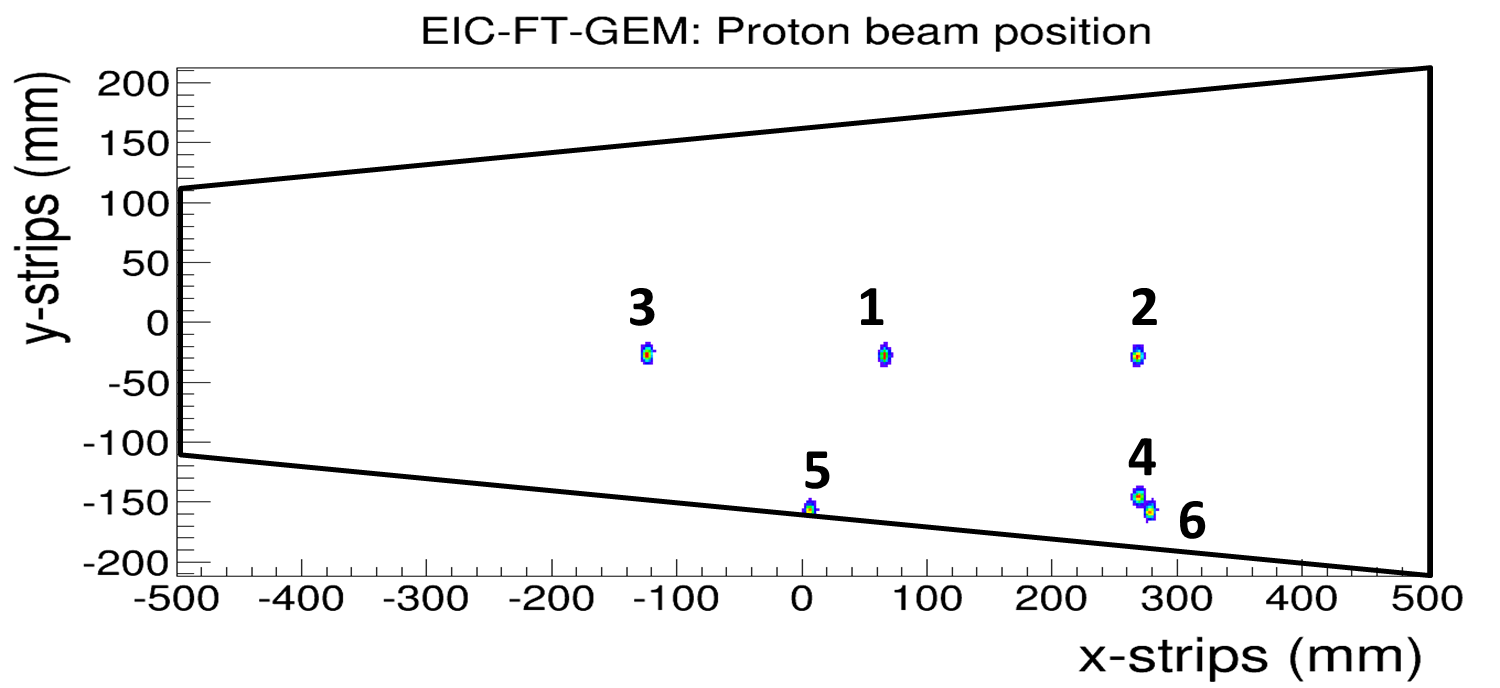}
\caption{\label{fig:eic_protonBeamScan} 2D beam profile from reconstructed particle positions}
\end{subfigure}
\begin{subfigure}[t]{0.35\textwidth}
\includegraphics[width=0.95\textwidth,natwidth=1003,natheight=842]{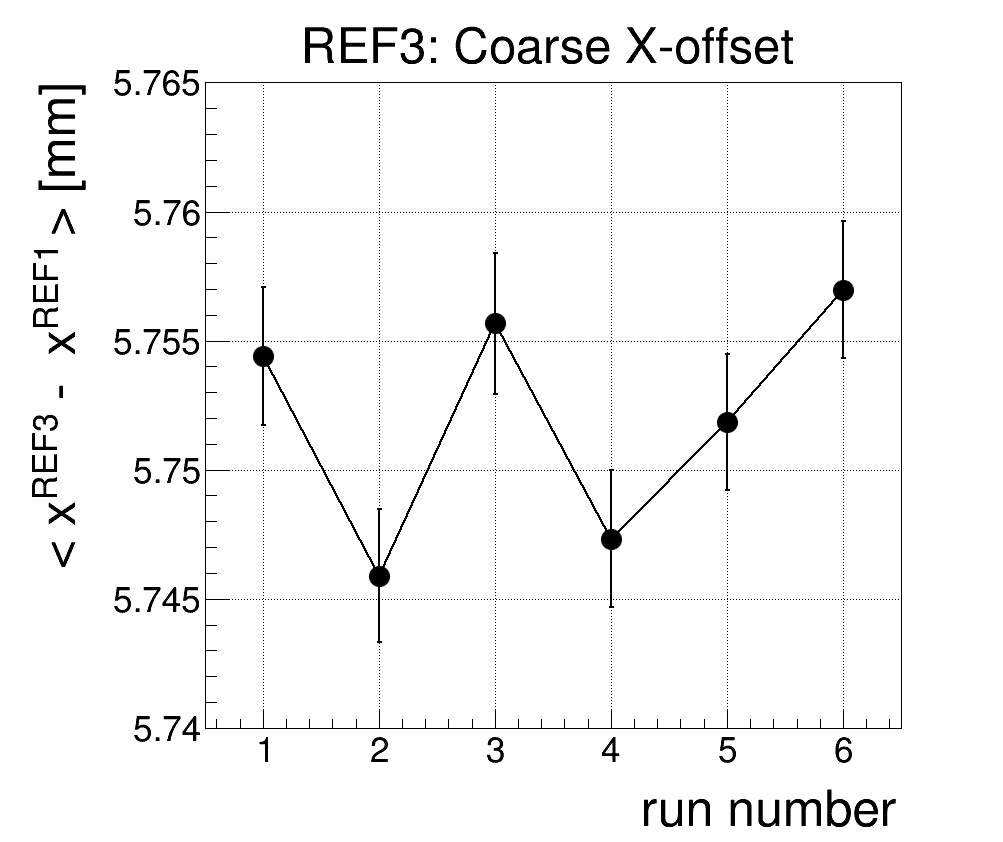}
\caption{\label{fig:offsetxBeamPos} Offset of REF3 in x-axis as a function of run number}
\end{subfigure}
\caption{\label{fig:protonBeamScan} Position scan at 6 locations on the chamber with 120 GeV proton beam at FTBF}
\end{figure}

The offset parameters for each detector is the mean value of a Gaussian fit to the distribution of $\left(x_{i}^{EIC} - x_{i}^{REF1}\right)$ and $\left(y_{i}^{EIC} - y_{i}^{REF1}\right)$ over all triggered event $i$, with  $\left(x^{EIC}_{i},y^{EIC}_{i}\right)$ representing the particle position coordinates measured in EIC-FT-GEM chamber. The correction X-offset is shown for tracker REF3 in Fig. \ref{fig:offsetxBeamPos} in x-direction as a function of the run number. The run number (horizontal axis) represents the position of the beam spot during the scan of the chamber, same as the number shown in  Fig. \ref{fig:eic_protonBeamScan}. The variation of X-offset from one run to another is as high as 15 $\mu$m and larger than the statistical error and could be explained by experimental conditions. Several GEM chambers with various sizes and types were positionned on the moving table alongside EIC-FT-GEM chamber as mentionned in Section \ref{sec:ftbfsetup}. As a consequence, the number of detectors and subsequently the amount of material on the path of the proton beam varies significantly from one run to another. We believe that this accounts for the small but significant variation of the alignment parameters from run to run.
\subsubsection{Correction of the x-y plane rotation}
\begin{figure}[!ht]
\vspace{-0.0cm}
\captionsetup{width=0.95\textwidth}
\centering
\begin{subfigure}[t]{0.485\textwidth}
\includegraphics[width=0.95\textwidth,natwidth=1050,natheight=930]{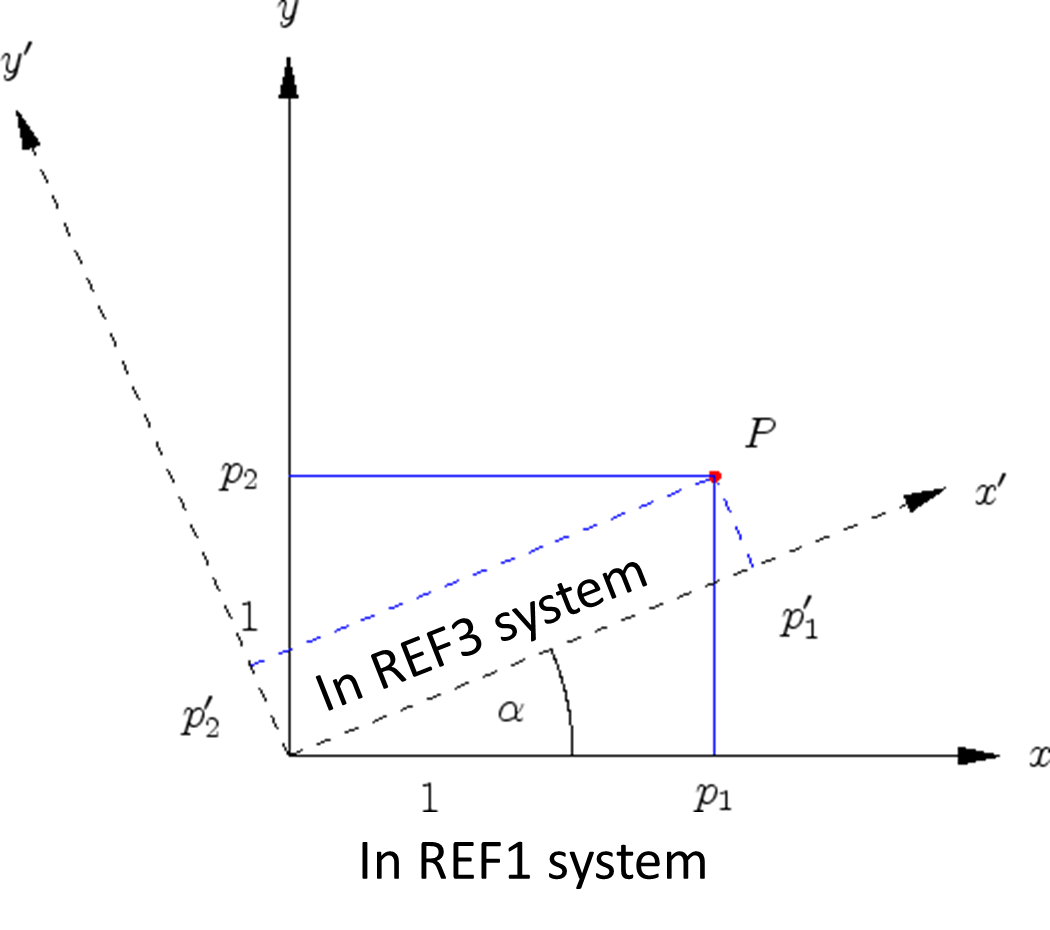}
\caption{\label{fig:planerotationmethod} plane rotation correction method}
\end{subfigure}
\begin{subfigure}[t]{0.485\textwidth}
\includegraphics[width=0.95\textwidth,natwidth=1003,natheight=842]{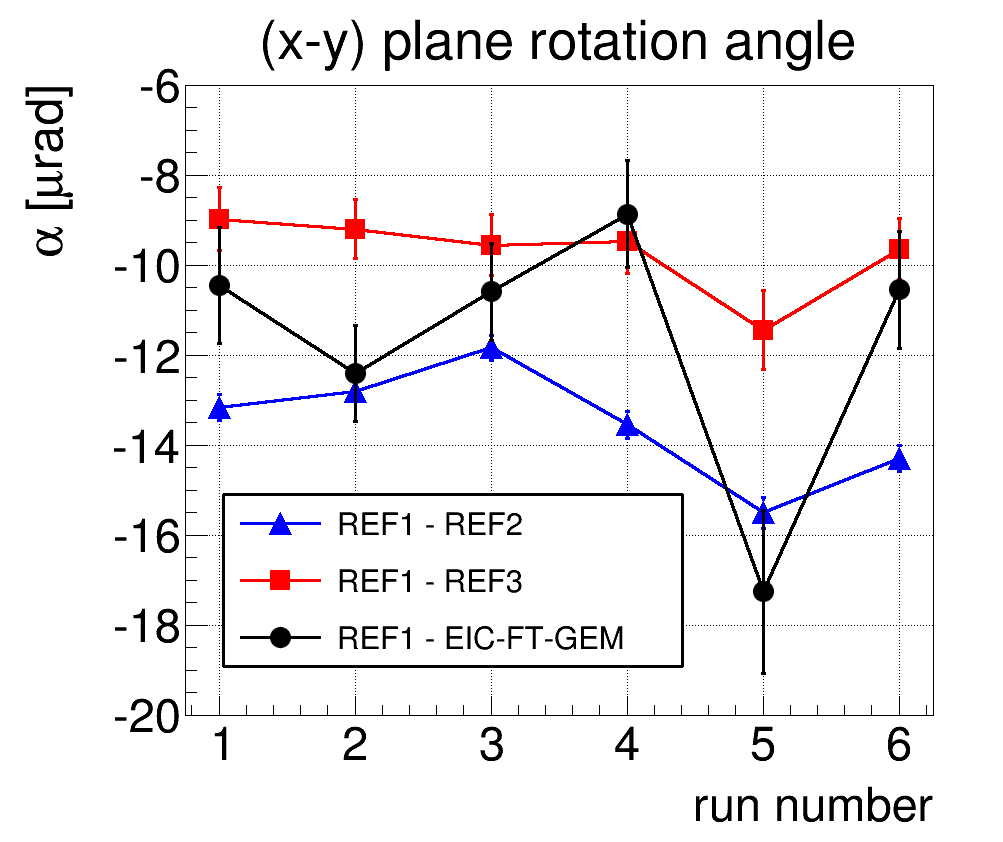}
\caption{\label{fig:planeRotationBeamPos} Rotation angle as a function of the run number}
\end{subfigure}
\caption{\label{fig:rotationParameters} Correction of the x-y plane rotation}
\end{figure}

The  correction of the offsets is followed by the correction of the rotation of the detectors x-y plane around the beam axis (z). The principle of the plane rotation is illustrated in Fig. \ref{fig:planerotationmethod}. The rotation angle $\alpha$ is defined as the mean of a Gaussian fit to the distribution of the $\alpha_{i}$ for each event as shown in [Eq. \ref{eq:rotAngle}]. Fig. \ref{fig:planeRotationBeamPos}  shows $\alpha$ for the reference trackers REF2, REF3 and EIC-FT-GEM chamber as a function of the scan run number. Once again, the variation of the rotation angle for different  runs is caused by the different amount of detector material on the beam path. 
\begin{equation} \label{eq:rotAngle}
\alpha_{i} = \arcsin  \left( \left( x_{i}' y_{i} -  y_{i}' x_{i} \right) / \left( x_{i}^2 + y_{i}^2 \right) \right)
\end{equation}
\subsubsection{Residuals}
\begin{figure}[!ht]
\vspace{-0.cm}
\captionsetup{width=0.9\textwidth}
\centering
\includegraphics[width=0.97\textwidth,natwidth=1448,natheight=910]{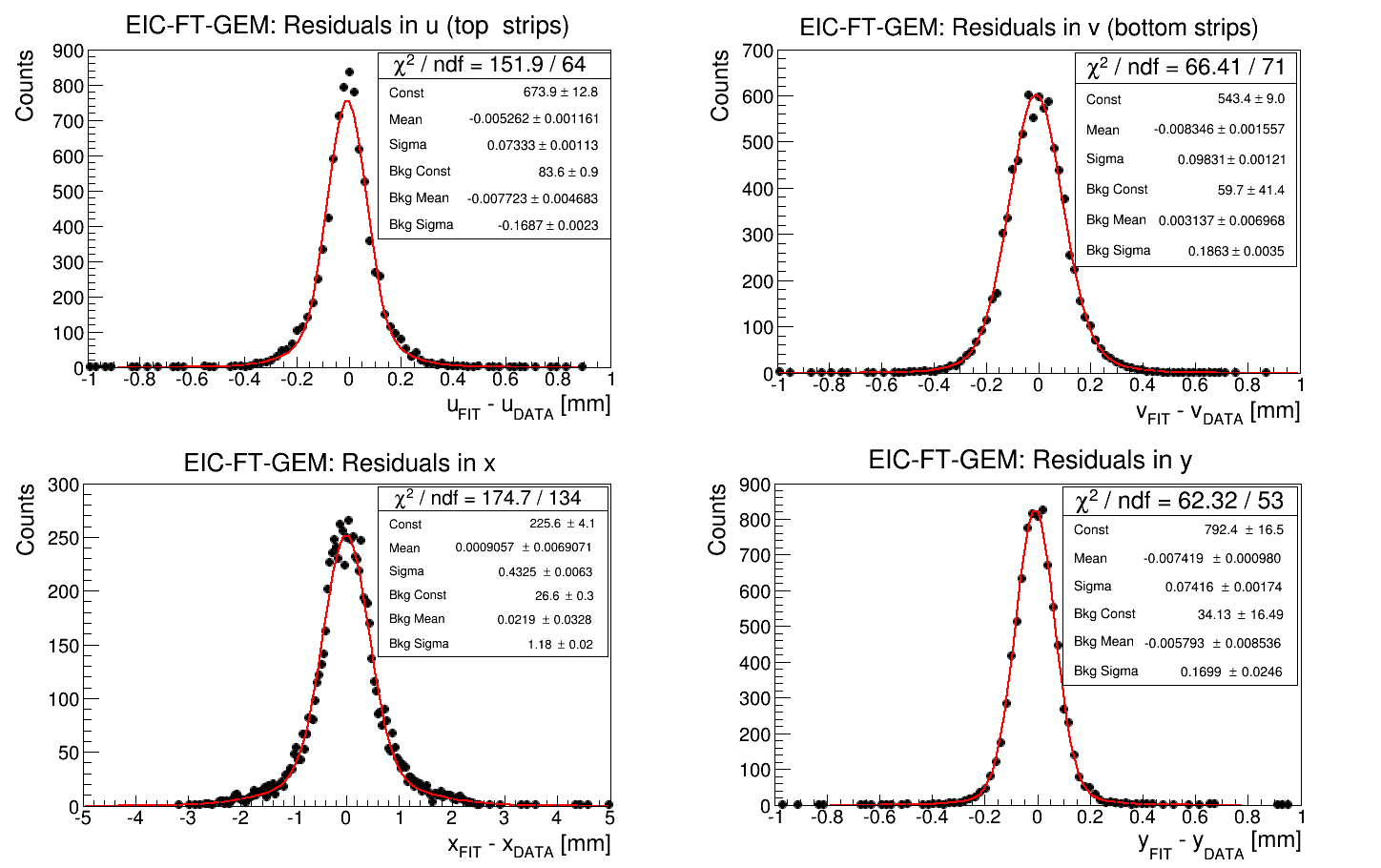}
\caption{\label{fig:residuals} tracks residuals of the particle position coordinates along u and v strips (top) and in x and y (bottom) }
\end{figure}

A simple least-squares fit algorithm is used to fit tracks from the measured coordinates from the reference trackers. One small GEM (REF1) and both large SBS GEMs (REF1, REF2)are used as reference trackers as shown in Fig. \ref{fig:ftbfSetup}. The other two small GEMs were left out to minimize the impact of the multiple scattering on the track fit. The fit is performed for every event and the residual, defined as the difference between the measured coordinates on EIC-FT-GEM chamber and the expected coordinates from the track fit, is filled into histogram. The  distributions  of the residuals are produced for the Cartesian (x,y) and cylindrical (r,$\varphi$) coordinate systems as well as for (u,v) coordinates along the top and bottom layer strips. The coordinates $u_{fit}$ and $v_{fit}$ are extracted from the fitted coordinates $x_{fit}$ and $y_{fit}$ using the [Eq. \ref{eq:uvCoord}] in a reverse operation with respect to  [Eq. \ref{eq:cartesianCoord}]. The plots in Fig. \ref{fig:residuals} show the residuals distribution for u and v coordinates (top plots) and for x and y coordinates (bottom plots) for data corresponding to beam position scan run 2. 
\begin{align} \label{eq:uvCoord}
&u_{fit} = y_{fit} + (0.5 \times L - x_{fit}) \times \tan ( \alpha / 2 ) \nonumber\\
&v_{fit} = y_{fit} - (0.5 \times L - x_{fit}) \times \tan ( \alpha / 2 )
\end{align}
\subsection{Spatial Resolution}
\subsubsection{Track fit error estimates}
We performed a Monte Carlo simulation of the FTBF setup to generate the tracker data to estimate the intrinsic error from the track fit. For the simulation, the spatial resolution of trackers for  x and y  were obtained by smearing the coordinates with a Gaussian distribution of a width of 60 $\mu$m and assuming perfect resolution  (i.e $\sigma_{EIC}$ = 0) for EIC-FT-GEM chamber. In this configuration, the width of the residual for the EIC-FT-GEM chamber represents the track fit error estimate (EE) and is equal to 35 $\mu$m for the y-coordinate and 40 $\mu$m and 45 $\mu$m for u and v coordinates respectively. The error is then recalculated for x-coordinate from the U-V strip readout and is equal 180 $\mu$m. The resolution  $\sigma_{RES}$ for each coordinate is obtained by subtracting in quadrature the track fit error (EE) from the width $\sigma_{res}$ of the residual distribution of the coordinate as shown in Eq. (\ref{eq:resolution}). 
\begin{equation}  \label{eq:resolution}
\sigma_{RES} = \sqrt{ \left( \sigma^2_{res} - \sigma^2_{EE} \right)}
\end{equation}
\subsubsection{Resolutions}
\begin{figure}[!ht]
\vspace{-0.0cm}
\captionsetup{width=0.9\textwidth}
\centering
\begin{subfigure}[t]{0.485\textwidth}
\includegraphics[width=0.95\textwidth,natwidth=796,natheight=572]{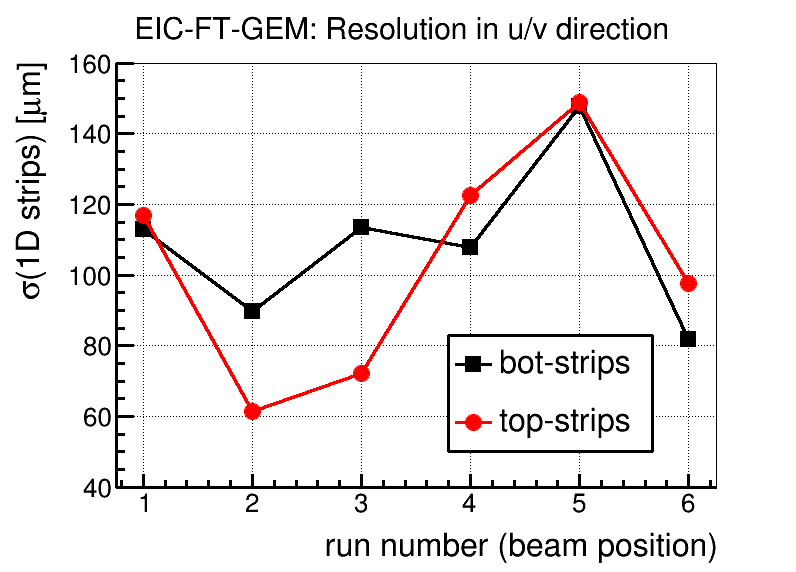}
\caption{\label{fig:uvResolution} Resolution of the coordinates along u and v strips.}
\end{subfigure}
\begin{subfigure}[t]{0.485\textwidth}
\includegraphics[width=0.95\textwidth,natwidth=796,natheight=572]{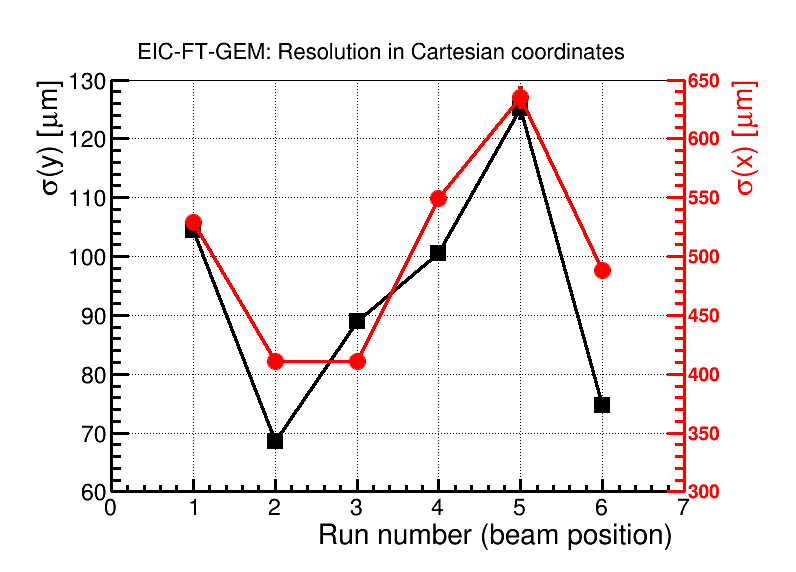}
\caption{\label{fig:xyResolution} Resolution in Cartesian coordinate system (x,y).}
\end{subfigure}
\caption{\label{fig:Resolution} Spatial resolution in six locations (identified by run number) on EIC chamber with proton beam}
\end{figure}

The spatial resolution along (u,v) directions  measured  on top and bottom layer strips are shown on the left plot in Fig. \ref{fig:uvResolution} as a function of the beam position (run number). Better position resolution performances are in principle expected for the top layer strips because of the larger cluster size and larger collection of charge (ADC counts) after zero suppression (as discussed in Fig. \ref{fig:clusterSize} of Section \ref{sec:clusterSize}). However, due to the complex structure of EIC chamber readout board with various strip lengths at different location of the active area (see Section \ref{sec:pedestal}), we  measured a better spatial resolution on the bottom strips for the beam at the position corresponding to the run 1, 4 and 6. Overall, the resolution is better than 120 $\mu$m  for both bottom and top layer strips for all locations except for run 5. This run corresponds to the proton beam hitting at the edge of the detector (see Fig. \ref{fig:eic_protonBeamScan}) leading to truncated clusters for both top and bottom strips and therefore affecting the measured position resolution. 
On the right, Fig. \ref{fig:xyResolution} shows the position resolution for  x and y coordinates. In the y-direction, the average resolution $\sigma(y)$ in over the five locations is better than 90 $\mu$m. However, variation is observed at different locations as expected, with the best resolution measured at beam position 2 equal to 70 $\mu$m and the worst (excluding location 5) at beam position 1 equal to 105 $\mu$m. In the x-direction, the measured resolution varies from 410 $\mu$m  to 550 $\mu$m. The average resolution in x is equal to $\sigma(x)$ = 480 $\mu$m  and about 5 times higher than $\sigma(y)$ = 90 $\mu$m as expected for a 12\degree angle of the 2D U-V strip readout board.
\begin{figure}[!ht]
\vspace{-0cm}
\captionsetup{width=0.9\textwidth}
\centering
\begin{subfigure}[t]{0.485\textwidth}
\includegraphics[width=0.95\textwidth,natwidth=1469,natheight=981]{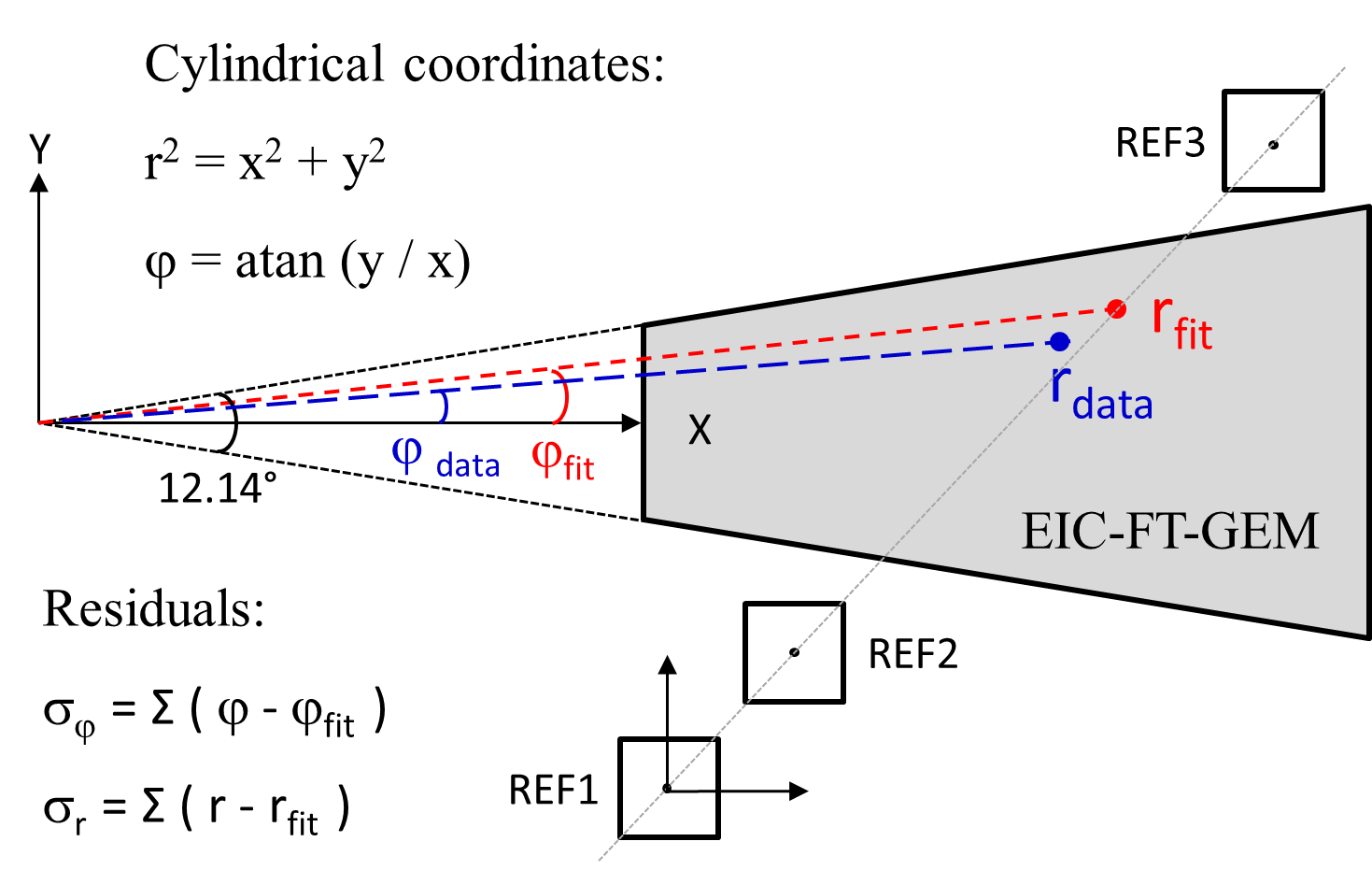}
\caption{\label{fig:cylindricalCoordSys} Transformation from Cartesian to Cylindrical coordinate.}
\end{subfigure}
\begin{subfigure}[t]{0.485\textwidth}
\includegraphics[width=0.95\textwidth,natwidth=796,natheight=572]{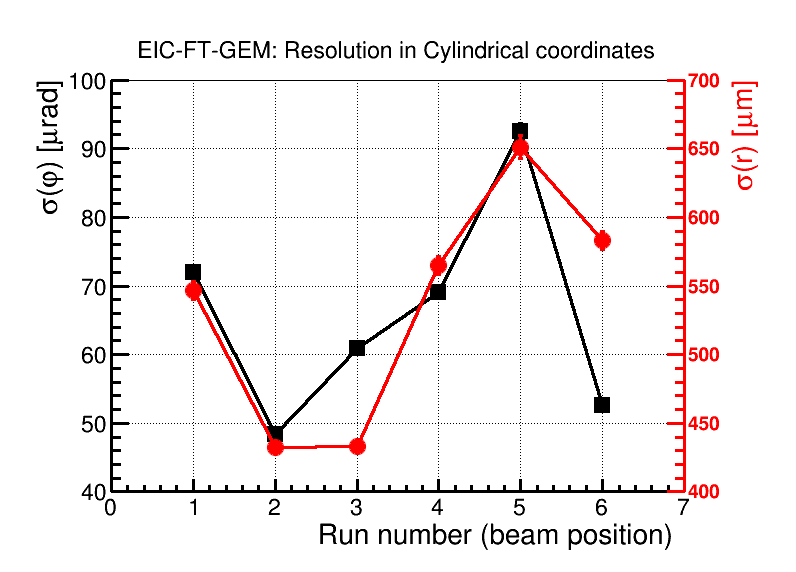}
\caption{\label{fig:rphyResolution} Resolution in Cylindrical coordinate (r,$\varphi$).}
\end{subfigure}
\caption{\label{fig:cylindricalResolution} Spatial resolution in Cylindrical coordinate system.}
\end{figure}
Cartesian coordinates (x,y)  could be easily transformed into (r, $\varphi$) in cylindrical coordinate system which is the natural coordinate system of an EIC forward tracker disk layer configuration. The  transformation has been applied to EIC-FT-GEM chamber. The vertex of the triangle formed by the radial sides of the trapezoidal chamber is used as origin for the transformation as illustrated in Fig. \ref{fig:cylindricalCoordSys}.  The resolution $\sigma(r)$ and $\sigma(\varphi)$ are shown in  Fig. \ref{fig:rphyResolution}. In the azimuthal direction, the average resolution of  $\sigma(\varphi)$ is equal to 60 $\mu$rad and in the radial direction, the average resolution  $\sigma(r)$ is  550 $\mu$m.  
\section{Conclusion} \label{sec:Conclusion}
A large-area and light-weight trapezoidal triple-GEM detector with a novel U-V strip readout on flexible PCB board has been assembled and tested. The performance of the chamber in test beam  shows an excellent detector response uniformity with an efficiency higher than 97\% across the active area. Spatial resolution better than 110 $\mu$m was achieved for both top and bottom strips in both the u and v direction with the 2D stereo-angle readout board. This translates to an angular resolution of 60 $\mu$rad  in the azimuthal direction and a position resolution of 550 $\mu$m in the radial direction. The excellent performances achieved for this first prototype demonstrates that the large-area and light-weight GEM with U-V strip readout would be a strong candidate for the large forward tracking detectors of the future experiments at the EIC.
\section*{Acknowledgements} 
This work is supported by the site neutral detector R$\&$D program administered by Brookhaven National Laboratory (BNL) for the Electron Ion collider (EIC) under the eRD6 consortium. We thank Rui de Oliveira at the PCB production facility, at CERN, for the technical support with the design and construction of the detector and the staff at the Fermilab Test Beam Facility for their friendly assistance during the test beam. We particularly want to thank Aiwu Zhang and Marcus Hohlmann (Florida Tech) with whom we share the large area detector setup during the test beam campaign at Fermilab and all our colleagues within the eRD6 consortium for the very valuable exchanges of ideas and suggestions during the analysis of the test beam data.

\end{document}